\documentclass[prd,aps,floats,floatfix,eqsecnum,nofootinbib]{revtex4}
\usepackage{verbatim,graphicx,amssymb,amsbsy,bm,amsmath,rotating,epsfig}
\usepackage{psfrag}
\usepackage{hyperref}
\usepackage{fontenc}
\usepackage[latin1]{inputenc}
\usepackage{pstricks,pst-node,pst-text,pst-3d}
\usepackage{textcomp}
\newcommand{\be}{\begin{equation}}
\newcommand{\ee}{\end{equation}}
\newcommand{\bea}{\begin{eqnarray}}
\newcommand{\eea}{\end{eqnarray}}

\newcommand{\br}{\ensuremath{\mathbf{r}}}

\begin{document}
\title{Warm Dark Matter Galaxies with Central Supermassive Black Holes}
\author{\bf H. J. de Vega $^{(+)}$}
\author{\bf N. G. Sanchez $^{(a)}$} 
\affiliation{$^{(+)}$ CNRS LPTHE Sorbonne
Universit\'e Universit\'e Pierre et Marie Curie
UPMC Paris, France \\
$^{(a)}$ CNRS LERMA PSL Observatoire de Paris
Sorbonne Universit\'e \\ and Chalonge - de Vega
International School Center, Paris, France.\\}
\date{\today}
\begin{abstract}
We generalize the Thomas-Fermi approach to galaxy
structure to include central supermassive black
holes and find selfconsistently and non-linearly
the gravitational potential of the galaxy plus the 
central black hole (BH) system. This approach naturally
incorporates the quantum pressure of the fermionic
warm dark matter (WDM) particles and  
shows its full power and clearness in the presence
of supermassive black holes. We find the main galaxy
and central black hole magnitudes as the halo radius $
r_h $, halo mass $ M_h $, black hole mass $M_{BH}$,
velocity dispersion $\sigma$, phase space density,
with their realistic astrophysical values, masses and
sizes over a wide galaxy range. The supermassive black
hole masses arise {\it naturally} in this framework. Our
extensive numerical calculations and detailed analytic
resolution of the Thomas-Fermi equations show that in
the presence of the central BH {\bf both} DM regimes:
classical (Boltzmann dilute) and quantum (compact) do
{\bf necessarily}  co-exist generically in {\bf any}
galaxy: from the smaller and compact galaxies to the
largest ones. The ratio $ {\cal R}(r)$ of the particle
wavelength to the average interparticle distance shows
consistently that the transition, $ {\cal R} \simeq 1
$, from the quantum to the classical region occurs
precisely at {\bf the same point} $ r_A $ where the
chemical potential vanishes. A {\bf novel halo
structure} with three regions shows up: In the
vicinity of the BH, WDM is {\bf always} quantum in a
small compact core of radius $r_A $ and nearly
constant density. In the region $ r_A < r < r_i$ till
the BH influence radius $r_i$, WDM is less compact and
exhibits a clear classical-Boltzmann like behaviour.
For $ r > r_i $, the WDM gravity potential dominates
and the known halo galaxy shows up with its
astrophysical size, DM is a dilute classical gas in
this region. As an illustration, three representative
families of galaxy plus central BH solutions are found
and analyzed: small, medium and large galaxies with
realistic supermassive BH masses of $10^5 M_\odot$ 
, $10^7 M_\odot$ and $10^9 M_\odot$ respectively. In
the presence of the central BH, we find a {\it
minimum} galaxy size and mass $ M_h^{min} \simeq  10^7
\;  M_\odot $, {\bf larger} 
($2.2233 \; 10^3$ times) than the one without BH, and
reached at a minimal {\it non-zero} temperature $
T_{min} $. The supermassive BH {\bf heats-up} the DM
and prevents it to become an exactly degenerate gas at
zero temperature. Colder galaxies are smaller, warmer
galaxies are larger. Galaxies with a central
black hole have large masses $ M_h > 10^7 \; M_\odot >
M_h^{min} $; compact or ultracompact dwarf galaxies in
the range $ 10^4 M_\odot < M_h < 10^7 M_\odot
$ {\bf cannot} harbor central BHs. We find {\bf novel}
scaling relations $M_{BH} = D M_h^\frac38 $ and  $ r_h
= C M_{BH}^\frac43 $, and show that the DM galaxy
scaling relations: $ M_h = b \; \Sigma_0 r_h^2$, $ M_h
= a\; {\sigma_h}^4 /\Sigma_0  $ hold too in the
presence of the central BH, $\Sigma_0$ being the
constant surface density scale over a wide galaxy
range.
The galaxy equation of state is derived: The pressure
$ P(r)$ takes huge values in the BH vecinity region
and then sharply decreases entering the classical
region following consistently a self-gravitating
perfect gas $P(r) = {\sigma}^2 \rho (r) $  behaviour.

\bigskip

 (+) : passed away : https://chalonge-devega.fr/HdeV.html \\
(a): https://chalonge-devega.fr/sanchez/ \\     Norma.Sanchez@obspm.fr                
\end{abstract}
\pacs{95.35.+d, 98.52.-b, 98.52.Wz}
\keywords{Dark Matter, Galaxy structure, Galaxy Density Profiles, Supermassive black holes}
\maketitle
\tableofcontents

\section{Introduction and Results}

Dark matter (DM) is the main component of galaxies:
the fraction of DM over the total
galaxy mass goes from 90\% for large diluted galaxies
till 99.99\% for dwarf compact galaxies.
Therefore, as a first approximation,  DM alone should
explain the {\bf main basic magnitudes} of galaxies
(as masses and sizes) as well as main structural
properties of density profiles and rotation curves.
Baryons should  give corrections to the pure DM
results. For such reasons we consider here 
Warm Dark Matter galaxies with central supermassive
black holes without including
baryons as a first approximation.

\medskip

Warm Dark Matter (WDM), that is dark matter formed by
particles with masses of the order of the keV scale
receives increasing attention in the last years, (see for example \cite{mnras},
\cite{cosmf}, \cite{highm}, \cite{highp}, \cite {cosmoprd}, \cite{evolwdm}, \cite{moreno},  \cite{boya}, \cite{newas}, \cite{astro}, \cite{urc},  \cite{eqesta}, \cite{eddi}, \cite{menci2016}, \cite{whitep},
\cite{prletters},  \cite {rudakov}, \cite{universe2021} and references therein).
At intermediate scales $ \sim 100 $ kpc, WDM gives the
{\bf correct abundance} of substructures 
and solves the cold dark matter (CDM) overabundance of
structures at small scales 
\cite{colin}, \cite{dolgov}, \cite{theuns},  \cite{tikho}, \cite{zav}, \cite{pap}, \cite {lov12},\cite{lovl}, \cite{ander}.
For scales larger than $ ~ 100 $ kpc, WDM yields the same results than CDM.
Hence,  WDM agrees with the 
small scale as well as large scale structure observations and CMB anisotropy observations.

{\vskip 0.1cm} 

Astronomical observations show that the DM galaxy density profiles are {\bf cored} 
till scales below the kpc \ \cite{obs},\cite{gil}, \cite{wp}, \cite{span}, \cite{dona}, \cite{kor}.
On the other hand, $N$-body CDM simulations exhibit cusped density profiles with a typical $ 1/r $ behaviour
near the galaxy center $ r = 0 $. 
Inside galaxy cores, below  $ \sim 100$ pc, $N$-body classical physics simulations 
do not provide the correct structures for WDM because quantum effects are important in WDM at these scales.
Classical that is non quantum physics $N$-body WDM simulations which do not take into account the quantum WDM pressure exhibit cusps or small 
cores with sizes smaller than the observed cores \citep{avi}, \cite{colin8}, \cite{vinas}, \cite{sinz}.
WDM predicts correct structures and cores with the right sizes for small scales (below kpc) 
when the {\bf quantum} nature of the WDM particles, that is the {\bf quantum pressure} of the fermionic WDM, is taken into account \cite{newas}, \cite{astro},\cite{urc},\cite{eqesta}.

\medskip

We follow here the Thomas-Fermi approach to galaxy structure for self-gravitating 
fermionic WDM \cite{newas}, \cite{astro}, \cite{urc}, \cite{eqesta} . This approach is especially appropriate to take into account 
quantum properties
of systems with large number of particles, namely macroscopic quantum systems
as neutron stars, white dwarf stars \citep{ll} and galaxies \cite{newas}, \cite{astro},\cite{urc}, \cite{eqesta} . 

Fermionic dark matter is appropriate because Dark Matter particles  do not interact with the standard or electromagnetic forces, a typical example is the sterile neutrino. Fermionic statistics is totally valid for Dark Matter and the keV Fermionic Dark Matter becomes popular in the last years. The DM particles composing the self-gravitating Fermi gaz only interact through gravitation.

In this paper we generalize the Thomas-Fermi approach to galaxies including their central supermassive black holes.

In this approach, the central quantity to derive is the DM chemical potential $ \mu(\br) $,
which is the free energy per particle. For self-gravitating systems,
the potential $ \mu(\br) $ is proportional to the gravitational potential $ \phi(\br) $,
$ \mu(\br) =  \mu_0 - m \; \phi(\br) $, $ \mu_0 $ being a constant, and 
obeys the {\bf self-consistent} and {\bf nonlinear} Poisson equation
\be\label{poisI}
\nabla^2 \mu(\br) = -4 \; \pi \; g \; G \; m^2 \; 
\int \frac{d^3p}{(2 \, \pi \; \hbar)^3} \; f\left(\frac{p^2}{2 \, m}-\mu(\br)\right) \; .
\ee
Here $ G $ is Newton's gravitational constant,  $ g $ is the number of internal degrees of freedom 
of the DM particle, $m$ is the DM particle mass, $ p $ is the DM particle momentum and $ f(E) $ is the
energy distribution function. This is a semiclassical gravitational approach to determine selfconsistently the
gravitational potential of the quantum fermionic WDM given its distribution function $ f(E) $.

\medskip

In the Thomas-Fermi approach, DM dominated galaxies are considered in a stationary state.
This is a realistic situation for the late stages of structure formation
since the free-fall (Jeans) time $ t_{ff} $ for galaxies is much shorter than the age of galaxies.
$ t_{ff} $ is at least one or two orders of magnitude smaller than the age of the galaxy.

\medskip

We consider spherical symmetric configurations where eq.(\ref{poisI}) becomes an
ordinary nonlinear differential equation that determines self-consistently
the chemical potential $ \mu(r) $ and constitutes the Thomas--Fermi approach \citep{newas},  \cite{astro}, \cite{eqesta},\cite{urc}.
We choose for the energy distribution function a Fermi--Dirac distribution
$$
f(E) = \frac1{e^{E/T_0} + 1} \; ,
$$
where $ T_0 $ is the characteristic one--particle energy scale. 
As we see below, except near the central black-hole we can take $ T_0 $ constant.
$ T_0 $ plays the role of an effective temperature scale and depends on the galaxy mass \citep{urc}, \cite{eqesta}. 

{\vskip 0.1cm} 

The Fermi--Dirac distribution function
is justified in the inner regions of the galaxy,
inside the halo radius where we find that the Thomas--Fermi density
profiles perfectly agree with the observations \citep{urc}, \cite{eqesta}. 
These results are supported by our work Ref.\cite{eddi} where 
within an Eddington-like approach for galaxies, it is shown that the observed galaxy density profiles
describe a self-gravitating thermal gas for $ r \lesssim R_{virial} $.

\medskip

Our theoretical results follow by solving the self-consistent and nonlinear Poisson equation 
eq.(\ref{poisI}) which is {\bf solely} derived from the purely {\bf gravitational} interaction
of the WDM particles and their {\bf fermionic} nature.

\medskip

The central quantity in the Thomas--Fermi equations (\ref{poisI}) is the chemical potential $ \mu(r) $.
The boundary condition of the chemical or gravitational potential $ \mu(r) $ at the center $ r \to 0$ in the Thomas-Fermi approach is extended here to allow for the presence of the central black hole, namely,
\be\label{MUBH}
\mu(r) \buildrel_{r \to 0}\over= \frac{G \; m \; M_{BH}}{r} + Const. + {\cal O}(r) \; .
\ee
$ M_{BH} $ being the black hole mass. That is, the presence of a galactic central black-hole implies near the center $ r \to 0 $, a behavior proportional to $ 1/r $ in  $ \mu(r) $, while in the absence of the black-hole 
let us recall that $ \mu(r) $ is bounded for $ r \to 0 $ \cite{newas}, \cite{astro}, \cite{urc} ,\cite{eqesta}.
 
Positive values of $ \mu(r) $ correspond to a self-gravitating quantum gas regime while negative values of
$ \mu(r) $ describe the self-gravitating classical (Boltzmann) regime \citep{ll}. As we see below, one of the results of this paper is that in galaxies possesing central black-holes, {\bf both} regimes do appear.
The strong gravitational field of the central black hole makes the WDM chemical potential large and positive
near the center. This implies that the WDM behaves {\bf quantum mechanically} inside a small {\bf quantum core}
with a nearly constant density.

\medskip

We summarize in what follows the main results of this paper:

\medskip

\begin{itemize}

\item {\bf (i)} We find  that $ \mu(r) $ takes large positive values in the inner regions as implied by eq.(\ref{NUBH}), 
then decreases till vanishing at $ r = r_A $,  and becomes negative for $ r > r_A $,
as shown by our detailed resolution of the Thomas-Fermi equation [sec. \ref{3eje} and Fig. \ref{fnu}].
Therefore, $ r_A $ is precisely the transition between the quantum and classical DM behaviours, $ r_A $
plays the role of the {\bf quantum DM radius} of the galaxy
for galaxies exhibiting a central black hole.
Namely, inside $ r_A $ the WDM gas is a self-gravitating {\bf quantum} gas, while
for $ r \gtrsim r_A $ the WDM gas is a self-gravitating classical Boltzmann gas.
The size $ r_A $ of the quantum  WDM core turns to be smaller for increasing galaxy masses
and black hole masses. WDM inside a small core of radius $ r_A $ is in a quantum gas high density state, namely a Fermi nearly degenerate state with  nearly constant
density  $ \rho_A $. For the three representative families of galaxy solutions we find here, the values of $ r_A $ and $ \rho_A $
are given by eqs.(\ref{res3gal})-(\ref{grangal}). The density $ \rho_A $
is orders of magnitude larger than its values for $ r > r_A $ where the WDM is in the classical Boltzmann regime.
$ r_A $ runs between 0.07 pc to 1.90 pc for galaxies with virial masses
from $ 10^{16} \; M_\odot $ to $ 10^7 \; M_\odot $ [as shown in sec. \ref{3eje}].
In any case, $ r_A $ is much {\bf larger} than the Schwarzchild radius of the central black hole
which runs from $ 10^{-4} $ pc to $ 10^{-8} $ pc.

\medskip

This is an {\bf important result}: in the vicinity of the central black hole
the fermionic WDM is always in a quantum regime while far from the central black hole
the WDM follows a classical Boltzmann regime \cite{eqesta}. This is natural to understand: 
the strong attractive gravitational force near the central BH compacts the WDM 
and its high density makes it to behave quantum mechanically. On the contrary, far from the BH the gravitational forces are weak,
the WDM is diluted and it is then described by a classical Boltzmann gas.
Ultracompact dwarf galaxies also exhibit WDM in a quantum regime \cite{newas}, \cite{astro}, \cite{eqesta}.

\medskip 

\item {\bf (ii)} In addition, the black
hole has an influence radius $ r_i $. In the vicinity
of the black hole, the gravitational force due to the
black hole is larger than the gravitational force
exerced by the dark matter. 
The influence radius of the black-hole $ r_i $  is
defined as the radius where both forces
are of equal strength. Both forces point inwards and
always sum up. $ r_i $ turns to be larger than the
radius $ r_A $ where the chemical potential vanishes,
$ r_i > r_A $. The region $ r_A < r < r_i $ is
dominated by the central black hole and the WDM
exhibits there a classical behaviour. For $ r \lesssim
r_i $, we see from Figs. \ref{fnu}-\ref{nup} that both
$ \mu(r) $ and $ |d\mu(r)/dr| $, [or equivalently, the
dimensionless  potential $ \nu(\xi) $ and its
derivative $ |d\nu(\xi)/dx|$, $ x $= ln $r/r_h$],
follow the behaviour dictated by the central
black hole eq.(\ref{conbh}) which produce straight
lines on the left part of the logarithmic plots Figs.
\ref{fnu}-\ref{nup}. Consistently, for $ r \gtrsim r_i
, \; \nu(\xi) $ and $ |d\nu(\xi)/dx| $ are dominated
by the WDM and exhibit a similar behaviour to that of
the Thomas-Fermi solutions without a central
black hole \cite{newas}, \cite{astro}, \cite{urc}, \cite{eqesta}. \\
 Fig. \ref{rho} shows that the local density behaviour
 is dominated by the black hole
for $ r \lesssim r_i $. For $ r_i \lesssim r \lesssim
r_h $ the WDM gravitational field dominates over the black  hole field and the galaxy core shows up. 
For medium and large galaxies the core is seen as a
plateau. At the same time the chemical potential is
negative for $ r \gtrsim r_i > r_A $ and the WDM is a
classical Boltzmann gas in this region.

\medskip  

The surface density 
\be\label{densuI}
\Sigma_0 \equiv  r_h  \; \rho_0  \simeq 120 \; M_\odot
/{\rm pc}^2 \quad 
{\rm up ~ to } \; 10\% - 20\% \; ,
\ee
has the remarkable property of being nearly {\bf
constant} and independent of  luminosity in different
galactic systems (spirals, dwarf irregular and 
spheroidals, ellipticals) spanning over $14$
magnitudes in luminosity and over different 
Hubble types \citep{dona},  \cite{span}. It is therefore a
useful physical characteristic scale in terms of which express galaxy magnitudes.
 
\medskip

\item {\bf (iii)} We find the main galaxy
magnitudes as the halo radius $ r_h $, halo mass $ M_h
$, black hole mass $ M_{BH} $, velocity dispersion,
circular velocity, density, pressure and phase space 
density.  Analytic formulae are derived for them and
expressed in terms of the reference surface density  $
\Sigma_0 $.  Moreover, we can express the black hole
mass
as
\be\label{MBHxi0}
 M_{BH} = 2.73116 \; 10^4 \; M_\odot \;
 \frac{\xi_0}{\left[\xi_h \; I_2(\nu_0)\right]^{\frac35}}
 \;
\left(\frac{\Sigma_0 \;  {\rm pc}^2}{120 \;
M_\odot}\right)^{\! \! \frac35} 
 \; \left(\frac{2 \, {\rm keV}}{m}\right)^{\! \!
 \frac{16}5} \; 
\ee
 $ \xi_0$ being the dimensionless central radius and
 $I_2(\nu_0)$ the 2nd momentum of the distribution function
eq.(\ref{dfI}).
The black hole mass $ M_{BH} $ {\it grows} when $ \xi_0 $
grows. 
Notice that $ M_{BH} $ does not simply grow linearly with $
\xi_0 $ due to the presence 
of the factor $\left[\xi_h \; I_2(\nu_0)\right]^{-\frac35}
$.  

\medskip 

\item {\bf (iv) } We find in this approach
explicit realistic galaxy solutions with central
supermassive black holes and analyze three representative
families of them: Small size (mass) galaxies, intermediate
size (mass) galaxies, large size (mass) galaxies.

For a fixed value of the surface density $ \Sigma_0 $, the 
solutions are parametrized by two truly physical
parameters: the dimensionless central radius $ \xi_0 $ and the constant $ A $
characteristic of the chemical potential behaviour eq.(\ref{conbh}) at the center $\xi \to 0 $. The dimensionless central radius $ \xi_0 $  is explicitated in Eq.(\ref{NUBH}): This is the ratio of the relevant physical parameters $(m, M_{BH}, T)$ which appear in the chemical potential at the center. The constant $A$ is truly physical too and characterizes the boundary condition of the chemical potential at the center in the presence of the central supermassive black hole, Eq.(\ref{conbh}). In the absence of the central SMBH: $\xi_0 = 0$,  and the boundary condition at the center without BH: $ \nu (0) = A$  is recovered. 

We derive an illuminating expression for the central radius
$r_0$  for large galaxies $ M_h \gtrsim 10^6 \; M_\odot $
explicitely in terms of the black hole mass $M_{BH}$, the
halo mass $M_h$ and the reference surface density $
\Sigma_0 $. It follows from eqs.(\ref{E0}), (\ref{xi0num})
and (\ref{xihI}) that,
\be
r_0 = l_0 \; \xi_0 = 126.762 \; \sqrt{\frac{10^6 \;
M_\odot}{M_h}} \; \frac{M_{BH}}{10^6 \; M_\odot} \; 
\sqrt{\frac{120 \; M_\odot}{\Sigma_0 \; {\rm pc}^2}} \; \; {\rm pc} 
\ee

\item {\bf (v)} We find from our extensive
numerical calculations that the halo is thermalized at the
uniform
temperature $ T_0 $ and matches the circular temperature $
T_c(r) $ by $ r \sim 3 \; r_h $.
This picture is similar to the picture found in the absence
of the central black hole which follows from the observed
density profiles in the Eddington-like approach to galaxies
\cite{eddi}.
We obtain here in the Thomas-Fermi approach and in the presence of a central supermassive black hole that
the halo is thermalized at a uniform temperature $ T_0 $ inside $ r \lesssim 3 \; r_h $ which tends to 
the circular temperature  $ T_c (r) $ at $ r \sim 3 \; r_h
$ as illustrated in Fig. \ref{vel}. The circular
temperature is defined in terms of the circular
velocity as: $T_c(r) = \frac{m}3 \; v_c^2(r) $. 
The circular temperature  is discussed in  Section 3 .  We introduce the circular Temperature $T_c(r)$ in terms of the circular (virial) velocity  $v_c^2(r)$ in the same way the Temperature $T(r)$ is defined in terms of the velocity dispersion $T(r) = m v^2 (r) /3$.   The circular velocity $v_c^2(r)$ is  defined and found in Section 2. Near the central black hole, the space dependent temperature  $T_c (r)$ is given by equipartition and the virial theorem, as  shown by Eqs. (3.1), (3.2), (3.3).

\medskip

From our extensive numerical calculations we find that the
galaxy mass increases and the galaxy size
increases when the constant $ | A | $  characteristic of the 
the central behaviour of $ \nu(\xi) $ for $ \xi \to 0 $
eq.(\ref{conbh}) increases.
This is similar to the case in the absence of 
central black holes where $ A = \nu(0) $ \cite{newas}, \cite{astro},\cite{eqesta}. 

\item {\bf (vi)} {We plot in Fig. \ref{vel}
the circular velocity given by eq.(\ref{vtf}) 
vs. $ \log_{10} r/r_h $. For $ r> r_h $ the circular
velocity tends to the 
velocity dispersion as obtained from the Eddington equation
for realistic
density profiles \cite{eddi}. For $ r \to 0 $ the circular
velocity
grows as in eq.(\ref{vtf0}) due to the central black hole
field.}

\item  {\bf (vii)}   {We find in
eqs.(\ref{res3gal})-(\ref{grangal}) the WDM mass $ M_A $
inside the quantum galaxy radius  $ r_A $.
$ M_A $ represents only a small fraction of the halo or
virial mass of the galaxy
but it is a significant fraction of the black hole mass $
M_{BH} $. 
We see from eqs.(\ref{res3gal})-(\ref{grangal}) that $ M_A
$ amounts
for 20\% of  $ M_{BH} $ for the medium and large galaxies
and 45\% for the small galaxy.}

\item {\bf (viii)} We also measure the
classical and quantum gas character of the galaxy plus
black hole system by  
means of the ratio $ {\cal R}(r)$ between the particle de
Broglie wavelength and the average interparticle distance.
For $ {\cal R} \lesssim 1 $ the system is of classical
dilute nature while for $ {\cal R} \gtrsim 1 $ it is a
macroscopic quantum system. We find ${\cal R}(r)$ in terms
of the surface density and momenta of the gravitational or
chemical potential in dimensionless units $\nu (\xi)$ 
eqs.(\ref{defR}), (\ref{defRa}). Fig. \ref{R} shows $
\log_{10} \cal R $ vs. $ \log_{10} (r/r_h) $ for the three
representative galaxy solutions. The transition from the
quantum to the classical regime occurs precisely at {\bf
the same point} $ r_A $ where
the chemical potential vanishes (see Fig. \ref{fnu}), as it
must be, showing the consistency and powerful of our
treatment. This point defines the transition from the
quantum to the classical behaviour.

\item {\bf (ix)}  There is an {\bf important
qualitative} difference between galaxy solutions with a
black hole
($ \xi_0 > 0 $), and galaxy solutions without a black hole
($ \xi_0 = 0 $). In the absence of the central black hole,
the halo mass $ M_h $ reaches the minimal value $ M_h^{min}
$  Eq. (\ref{mhmin}) which is the degenerate quantum limit
at zero temperature $T_0^{min} = 0 $
\cite{newas}, \cite{astro}, \cite {eqesta}. In the presence of a central
black hole, we find that the minimal temperature $
T_0^{min} $ is always {\bf non-zero} and that the halo mass
takes as minimal value 
\be\label{bhmhmin}
 M_h^{min} =  6.892 \; 10^7 \; \left(\frac{2 \, {\rm keV}}{m}\right)^{\! \! \frac{16}5} \;
 \left(\frac{\Sigma_0 \;  {\rm pc}^2}{120 \;
 M_\odot}\right)^{\! \! \frac35}  \;  M_\odot 
\;, \; {\rm {\bf with ~ central ~ black ~ hole}} \; .
\ee

\be\label{mhmin}
 M_h^{min} = 3.0999 \; 10^4 \; \left(\frac{2 \, {\rm
 keV}}{m}\right)^{\! \! \frac{16}5} \;
 \left(\frac{\Sigma_0 \;  {\rm pc}^2}{120 \;
 M_\odot}\right)^{\! \! \frac35}  \;  M_\odot \;
, \;  T_0^{min} = 0 \;, \; {\rm {\bf
without ~ central ~ black ~ hole}} \; .
\ee

The presence of the supermassive black hole {\bf heats-up}
the dark matter gas
and prevents it to become an exact degenerate gas at zero
temperature. The minimal galaxy mass and size and most
compact galaxy state with a central black hole is a nearly
degenerate state at very low but non-zero temperature as
seen from eq.(\ref{bhtmin}). All matter studied in this paper is Dark Matter and the only DM interaction is the gravitational interaction. The presence of the black hole naturally makes the DM particles acquires a higher velocity (and thus a higher associated temperature), and in this sense the SMBH does "heat" the dark matter around it. Gravitation self-consistently, acts on such DM, and the SMBH adds too to such gravitational action.  This is a very clean physical process, clean framework and clean conclusive results.

This situation is clearly shown in Fig. \ref{amh}. The
value of $M_h^{min}$ with a central black hole is 
$2.2233 \; 10^3$ times larger than without the black hole.
Notice that the small galaxy solution eq.(\ref{res3gal}) is
just 11 \% larger in halo mass than the minimal galaxy
eq.(1.6) with central black hole. {\bf We conclude} that
galaxies possesing a central black hole are in the dilute
Boltzmann regime because of their large mass $ M_h > 10^6
\; M_\odot > M_h^{min} $ \cite{eqesta}. On the contrary,
compact galaxies, in particular ultracompact galaxies in
the quantum regime $ M_h < 2.3 \; 10^6  \;  M_\odot $
\cite{eqesta},
{\bf cannot} harbor central black holes because the minimal
galaxy mass with central black hole eq.(1.6) is always
larger than $ 2.3 \; 10^6  \;  M_\odot $. In other words,
galaxies with masses $ M_h < M_h^{min} $, namely
ultracompact dwarfs {\bf necessarily} do not possess
central black holes.

\medskip

The mass of the supermassive black hole $ M_{BH} $
monotonically increases with   the central radius  $r_0$ or equivalently the dimensionless one $\xi_0 $ at fixed $ A $.
In addition, for  $ \xi_0 < 0.3 $, that is for small
supermassive black holes,  and all $ A $, the galaxy
parameters as halo mass $ M_h $, halo radius $ r_h $,
virial mass $ M_{vir} $ and galaxy temperature $ T_0 $ 
become {\bf independent} of $ \xi_0 $ showing {\it a
limiting galaxy solution}. Only the BH mass
depends on  $ \xi_0 $ in this regime.

Fig. \ref{bhmh} displays our results for the black hole
mass $ \log_{10} M_{BH} $ vs. the halo mass $ \log_{10} M_h
$.
We see that $ M_{BH} $ is a {\bf two-valued} function of $
M_h $. For each value of $ M_h $ there are 
two possible values for $ M_{BH} $ which 
are quite close to each other. This two-valued dependence
on $ M_h $ is a direct consequence
of the dependence of $ M_h $ on $ A $ shown in Fig.
\ref{amh}.
The branch-points on the left in Fig.\ref{bhmh}, correspond
to the minimal galactic halo mass $ M_h^{min} $ 
eq.(\ref{bhmhmin}) when the central supermassive black hole
is present.
At {\bf fixed} $ \xi_0 $, as shown in Fig. \ref{bhmh}, the
central black hole mass $ M_{BH} $ {\bf scales} with the
halo mass $ M_h$ as
$$
 M_{BH} = D(\xi_0) \;  M_h^\frac38 \; ,
$$
where $ D(\xi_0) $ is an increasing function of $ \xi_0 $.
We plot in Fig. \ref{mht0} the halo galaxy mass $ \log_{10}
M_h $ vs. the galaxy temperature $ \log_{10} T_0/{\rm K} $.
The halo mass $ M_h $ grows when  $ T_0 $ increases. Colder
galaxies are smaller. Warmer galaxies are larger.
We see at the branch-points in Fig. \ref{mht0} the minimal
galaxy temperature $ T_0^{min} $
eq.(\ref{bhtmin}) when a supermassive black hole is
present.

We find galaxy solutions with central black holes for
arbitrarily small values of $ \xi_0 > 0 $ and
correspondingly arbitrarily small central BH mass. There is
no minimal central BH mass. The only minimal central BH
mass possibility is zero (for $\xi = 0$).

\item {\bf (x)} We find that $ M_h $ {\bf
scales} as $ r_h^2 $, which is the same scaling found in
the Thomas-Fermi approach to galaxies in the absence of
black holes \cite{newas}, \cite{astro}, \cite{eqesta}.
We plot in Fig. \ref{mhrh} the ordinary logarithm of the
halo radius $ \log_{10} r_h $ vs. 
the ordinary logarithm of the halo mass $ \log_{10} M_h $
for galaxies with
central black holes of many different masses. The halo mass
in the absence of a central black hole behaves in
the Thomas-Fermi approach as \cite{eqesta}
\be \label{bsinbh}
 M_h = 1.75572 \; \Sigma_0 \; r_h^2 \quad ,\quad {\rm {\bf
 without ~ central ~ black ~ hole}} \; .
\ee
The proportionality factor in this scaling relation is
confirmed by the galaxy data \cite{eqesta}.
In the presence of a central black hole we find in the
Thomas-Fermi approach 
an analogous relation
\be\label{bbh}
 M_h = b \; \; \Sigma_0 \; r_h^2 \quad, \quad {\rm {\bf
 with ~ central ~ black ~ hole}} \; .
\ee
where the coefficient $ b $ turns to be of order unity. We
plot in Fig.\ref{coef}
the coefficient $ b $ as a function of the halo mass $ M_h
$. We see that except for halo masses near the minimum
halo mass $ M_h^{min} $, $ b $ in the presence of a central
black hole takes values up to
10\% below its value in the absence of a central black hole
eq.(\ref{bsinbh}).
For halo masses near $ M_h^{min} , \; b $ increases reaching values $ b \leq 4 $.
For very large halos and central black holes, $b$ could be as small as about $1.6$.
$ b $ changes at most by a factor from $ 1/2 $ up to $ 2 $
while the halo mass $ M_h $ varies ten orders of magnitude.
As shown by Fig.\ref{coef}, $ b $ is a two-valued function
of $ M_h $. $ b $ turns to be independent of the precise
value of the WDM particle mass $ m $, which is due to the
fact that the scaling relation eq.(\ref{bbh}) as well as
eq.(\ref{bsinbh}) apply in the classical Boltzmann regime
of the galaxy ($ M_h \gtrsim 10^6 \; M_\odot $). {\bf In
summary}, the scaling relation 
eq.(\ref{bbh}) and the coefficient $ b $ turn out to be {\bf remarkably robust}.

\item {\bf (xi)} We plot in Fig. \ref{mbhrh}
the ordinary logarithm of the halo radius $ \log_{10} r_h $
versus 
the ordinary logarithm of the central black hole mass $
\log_{10} M_{BH} $ for many
galaxy solutions. The halo radius $ r_h $ turns to be a
double-valued function of $ M_{BH} $.
 Remarkably,  $ r_h $ for {\bf fixed} $ \xi_0 $ scales  as  
\be
r_h = C(\xi_0) \;  M_{BH}^\frac43 .
\ee
The constant $ C(\xi_0) $ turns to be a decreasing function of $ \xi_0 $.

\item {\bf (xii)}   We find the local pressure
$ P(r) $  as given by eq. (\ref{presion}). In Fig.\ref{pr}
we plot $ \log_{10} P(r) $ vs. $ \log_{10} (r/r_h) $ for
the three representative galaxy solutions.  $ P(r) $
monotonically decreases with $ r $. The pressure $ P(r) $
takes huge values in the quantum (high density) region $ r
< r_A $ and then it sharply decreases entering the
classical (dilute) region $ r > r_A $.
In Fig.\ref{prho}  we plot the derived {\bf equation of
state} $ \log_{10} P(r) $ vs. $ \log_{10} \rho(r)/ \rho_0 $
for the three galaxy solutions we find here with central
SMBH. 
The three curves almost coincide and are almost straight
lines of unit slope. That is, the equation of state is in
very good approximation a perfect gas equation of state
$P(r) = {\sigma}^2 \rho (r) $ , which
stems from the fact that galaxies with central black  holes
have halo masses
$ M_h > M_h \gtrsim 10^6 \; M_\odot  > M_h^{min} $
eq.(\ref{bhmhmin}) and therefore necessarily belong to the 
{\it dilute} Boltzmann classical regime \cite{eqesta}. The
equation of state turns out to be a local ($r$-dependent) perfect
gas equation of state because of the gravitational
interaction, (WDM self-gravitating perfect gas). 
Indeed, for galaxies with central black holes the WDM is in
a quantum (highly compact) regime 
inside the quantum radius $ r_A $. However, because $ r_A $
is in the parsec scale or smaller [see
eqs.(\ref{res3gal})-(\ref{grangal})] the bulk of the WDM is
in the Boltzmann classical regime which is consistently
reflected in the perfect gas equation of state behaviour.

\end{itemize}

In summary, the results of this paper show the power and
cleanliness of the Thomas-Fermi theory and WDM to properly
describe the galaxy structures and the galaxy physical
states with and without supermassive central black holes.
We consider in this paper pure WDM galaxies with central
supermassive black holes. Adding baryons will introduce
corrections, but the picture of galaxies with central
supermassive black holes presented here should not change
essentially.
This approach is {\bf independent} of any WDM particle
physics model. It depends only of the fermionic WDM nature
and gravity. The results presented in this paper do not
depend on the precise
value of the WDM particle mass $m$ but only on the fact
that $m$ is in the $keV$ scale, namely  keV $2 \lesssim m
\lesssim 10$ keV  say.

\medskip

This paper is organized as follows: 

\medskip

In Section II we formulate the problem of galaxy structure
with central supermassive black holes in the WDM
Thomas-Fermi approach and find the main physical magnitudes
and properties of the galaxy plus black hole system. In
Section III we solve the corresponding equations with the
boundary conditions, find three representative families of
galaxy solutions (small, medium and large size galaxies)
with central supermassive black holes and analyze the new
quantum and classical physics properties of the system. In
Section IV we perform an extensive study of the galaxy
solutions with a central supermassive black hole, find the
main important differences between galaxies with and
without the presence of a central black hole, derive
universal galaxy scaling relations in the presence of a
central supermassive black hole: halo mass $M_h$ - halo
size $r_h$ relation, black hole mass $M_{BH}$ - halo radius
$ r_h$ relation, and find the galaxy local pressure and
equation of state in the presence of central supermassive
black holes and their different regimes.  

\section{Galaxy structure with central supermassive
black holes in the WDM Thomas-Fermi approach}\label{fortoy}

We consider DM dominated galaxies in their late stages of structure formation when 
they are relaxing to a stationary situation, at least 
not too far from the galaxy center.

{\vskip 0.1cm} 

This is a realistic situation since the free-fall (Jeans) time $ t_{ff} $ for galaxies
is much shorter than the age of galaxies:
\be \label{tff}
t_{ff} = \frac1{\sqrt{G \; \rho_0}} = 1.49 \; 10^7 \; 
\sqrt{\frac{M_\odot}{\rho_0 \; {\rm pc}^3}} \; {\rm yr} \; .
\ee

The observed central densities of galaxies yield free-fall
times in the range
from 15 million years for ultracompact galaxies till 330
million years for
large diluted spiral galaxies. These free-fall (or
collapse) times are small compared
with the age of galaxies running in billions of years.

{\vskip 0.1cm} 

Hence, we can consider the DM described by a time
independent and 
energy distribution function $ f(E) $, where $ E =
\sqrt{p^2 + m^2} - m - \mu $
is the {\it relativistic} single-particle energy, $ m $ is
the mass of the DM particle and
$ \mu $ is the chemical potential \citep{newas}, \cite{astro}, \cite{eqesta}
related to the gravitational potential $ \phi(\br) $ by

\be \label{potq}
  \mu(\br) =  \mu_0 - m \, \phi(\br) \; ,
\ee

where $ \mu_0 $ is a constant. We consider here {\it relativistic} kinematics because the WDM
particles in the vicinity of the central black hole can be relativistic.
In the non-relativistic limit we recover the relations used in Refs.  \citep{newas}, \cite{astro}, \cite{eqesta}.

{\vskip 0.1cm} 

In the Thomas--Fermi approach, $ \rho(\br) $ is expressed
as a function of $ \mu(\br) $ through the
standard integral of the DM phase-space distribution
function over the momentum

\be \label{den}
  \rho(\br) = \frac{g}{2 \, \pi^2 \, \hbar^3} \int_0^{\infty} dp\;p^2 \; \sqrt{p^2 + m^2}
  \; f\left[\displaystyle \sqrt{p^2 + m^2} - m -\mu(\br)\right] \; , 
\ee

where $ g $ is the number of internal degrees of freedom of
the DM particle,
with $ g = 1 $ for Majorana fermions and $ g = 2 $ for
Dirac fermions. 
Eq.(\ref{den}) is valid in general for relativistic
fermions and {\it generalizes} the 
non-relativistic framework of refs. \cite{newas}, \cite{astro}, \cite{eqesta}.

\medskip

We will consider spherical symmetric configurations. Then, 
the Poisson equation for $ \phi(r) $ takes the
self-consistent form

\be \label{pois}
  \frac{d^2 \mu}{dr^2} + \frac2{r} \; \frac{d \mu}{dr} = - 4\pi \, G \, m \, \rho(r) =
- \frac{2 \, g \; G \; m}{\pi \; \hbar^3} \int_0^{\infty} dp\;p^2 \; \sqrt{p^2 + m^2}
  \; f\left[\displaystyle \sqrt{p^2 + m^2} - m -\mu(r)\right]\; , 
\ee

where $ G $ is Newton's constant and $ \rho(r) $ is the DM
mass density. 

\medskip

Eq. (\ref{pois}) provides an ordinary {\bf nonlinear}
differential equation that determines {\bf
self-consistently} the chemical potential $ \mu(r) $ and
constitutes the Thomas-Fermi approach \citep{newas}, \cite{astro}, \cite{eqesta}
(see also ref. \citep{peter}).
This is a semi-classical approach
to galaxy structure in which the quantum nature of the DM
particles is taken into account through
the quantum statistical distribution function $ f(E) $.

\medskip

The DM pressure and the velocity dispersion can also be
expressed as 
integrals over the DM phase-space distribution function as

\be\label{P}
 P(r) = \frac13 \; \rho(r) \; <v^2(r)> = 
\frac{g}{6 \, \pi^2 \; \hbar^3} \int_0^{\infty} dp\; \frac{p^4}{\sqrt{p^2 + m^2}}
\; f\left[\displaystyle \sqrt{p^2 + m^2} - m -\mu(r)\right] \;  , 
\ee

We see that $\mu(r)$ fully characterizes the DM halo
structure in this
Thomas-Fermi framework. 

{\vskip 0.1cm} 

In this semi-classical framework the stationary energy
distribution function $
f(E) $ must be given. We consider the Fermi-Dirac
distribution 
\be\label{FD}
  f(E) = \Psi_{\rm FD}(E/T_0) = \frac1{e^{E/T_0} + 1} \; ,
\ee
where the characteristic one-particle energy scale $ T_0 $
in the DM halo
plays the role of an effective temperature. 
$ T_0 $ can be taken constant except near the central
black hole.

{\vskip 0.1cm} 

In neutron stars, where the neutron mass is about six
orders of magnitude larger
than the WDM particle mass, the temperature can be
approximated by zero.

{\vskip 0.1cm} 

As shown in ref. \cite{eqesta}, the value of $ T_0 $ depends
on the galaxy mass.
In galaxies, $ T_0 \sim m \; <v^2> $ turns to be non-zero
but small in the range: 
$ 10^{-3} \; {\rm K} \lesssim T_0  \lesssim 10 $ K for all
halo galaxy masses in 
the range $ 10^5 - 10^{12} \; M_\odot $ which reproduce the
observed velocity
dispersions for $ m \simeq 2 $ keV. The smaller values of $
T_0 $ correspond to compact
dwarf galaxies and the larger values of $ T_0 $ are for
large and diluted galaxies \cite{eqesta}.

{\vskip 0.1cm} 

More precisely, large positive values of the chemical
potential 
correspond to the degenerate fermions limit which is the
extreme quantum state,  and oppositely, 
large negative values of the chemical potential at the
origin give the diluted states which are in  the classical regime. The quantum degenerate regime
 describes dwarf and compact galaxies while
the classical diluted regime describes large and diluted
galaxies.
In the classical regime, the Thomas-Fermi equation
(\ref{pois}) become the equations for a self-gravitating
Boltzmann gas.

{\vskip 0.1cm} 

Galaxies possesing central black holes exhibit {\it both} quantum and classical regions as we see below.

\medskip

The units used in this paper are those appropriate to the keV mass of the dark matter particle in the context of galaxy structure. The expression and conversion of units in terms of the keV includes the Planck constant h. Relevant conversion relations in terms of keV in this context are
\be
 keV \; kpc  =  1, 563738 \; 10^{29}
\ee
\be
M_\odot  =  1, 115468 \; 10^{63}\;keV
\ee

\subsection{Thomas-Fermi equations with a central black hole}

It is useful to introduce dimensionless variables $ \xi , \; \nu(\xi) $ 
\be\label{varsd}
 r = l_0 \; \xi \quad , \quad \mu(r) =  T_0 \;  \nu(\xi) \quad , \quad f(E) = \Psi(E/T_0) \; , 
\ee
where $ l_0 $ is the characteristic length that emerges from the dynamical equation (\ref{pois}):
\be\label{varsd2}
l_0 \equiv  \frac{\hbar}{\sqrt{8\,G}} \; \left(\frac2{g}\right)^{\! \! \frac13} \;
\left[\frac{9 \, \pi \; I_2(\nu_0)}{m^8\,\rho_0}\right]^{\! \! \frac16} 
= R_0 \; \left(\frac{2 \, {\rm keV}}{m}\right)^{\! \! \frac43}  \; \left(\frac2{g}\right)^{\! \! \frac13} \;
  \left[\frac{I_2(\nu_0)}{\rho_0} \; \frac{M_\odot}{{\rm pc}^3}\right]^{\! \! \frac16} 
 ,\; R_0 = 7.425 \; \rm pc  \; ,
\ee
and
\be\label{dfI}
I_2(\nu) \equiv 3 \; \int_0^{\infty} y^2 \; dy \; \sqrt{1 +\frac{2 \, y^2}{\tau}} \; 
\Psi_{FD}\left(\displaystyle \tau \left[\sqrt{1 +\frac{2 \, y^2}{\tau}}-1\right] -\nu\right) \;  ,
\ee
\be
\tau \equiv \frac{m}{T_0} \;  , \quad \nu_0 \equiv \nu(\xi_i) \; , \quad \rho_0 = \rho(\xi_i) 
\ee
where we use the integration variable $ y \equiv p / \sqrt{2 \, m \;  T_0} $.
$ \xi_i $ stands for the influence radius of the black hole which is defined below by
eq.(\ref{dnudxi}).

{\vskip 0.1cm} 

We consider in eq.(\ref{dfI}) the case of a constant temperature $ T_0 $.
The case of a $r$-dependent temperature is analysed in sec. \ref{tder}.

\medskip

For definiteness, we will take $ g = 2 $, Dirac fermions in the sequel. 
One can easily translate from Dirac to Majorana fermions changing the WDM fermion mass as:
$$
m \Rightarrow \frac{m}{2^\frac14} = 0.8409 \; m \; .
$$

\medskip

Then, in dimensionless variables, the self-consistent Thomas-Fermi equation 
(\ref{pois}) for the chemical potential $ \nu(\xi) $ takes the form

\be\label{nu}
\frac{d^2 \nu}{d\xi^2} + \frac2{\xi} \; \frac{d \nu}{d\xi} = - I_2(\nu) \quad .
\ee

The presence of the central black hole is introduced through the 
boundary conditions eq.(\ref{conbh}) given below for the chemical potential $ \nu(\xi) $ 
at $ \xi \to 0 $.

\subsection{Central Galactic Black Hole and its influence radius}

In the presence of a central galactic black hole the gravitational potential and the chemical potential
near the center take the form

\be\label{fimu0}
\phi(r) \buildrel_{r \to 0}\over= - \frac{G \; M_{BH}}{r} \quad ,  \quad \mu(r) -  \mu_0 \buildrel_{r \to 0}\over= \frac{G \; m \; M_{BH}}{r} \; ,
\ee

where  $ M_{BH} $ is the black hole mass. 
{\vskip 0.2cm} 

Integrating eq.(\ref{pois}) from $ r = 0 $ to $ r $ yields
\be\label{lim0}
r^2 \; \frac{d\mu}{dr} -\left[ \left. r^2 \; \frac{d\mu}{dr}\right|_{r \to 0} \right] = -G\,m\,M(r) \; 
\ee
where 
\be\label{dMr}
M(r) = 4\pi \int_0^r dr'\, r'^2 \, \rho(r') \; .
\ee
is the total WDM mass $ M(r) $ enclosed in a sphere of radius $ r $ not including the central black hole mass.

{\vskip 0.2cm} 

Inserting the $ r \to 0 $ behaviour eq. (\ref{fimu0}) into eq. (\ref{lim0}) yields for the derivative
of the chemical potential
\be\label{dmu}
\frac{d\mu}{dr} = -\frac{G \; m}{r^2} \; \left[M(r) + M_{BH} \right] \; , 
\ee
showing that the chemical potential is monotonically decreasing in $ r $.

{\vskip 0.2cm} 

From eqs.(\ref{potq}) and (\ref{varsd})
the dimensionless chemical potential $ \nu(\xi) $ takes the form
\be\label{NUBH}
\nu(\xi) \buildrel_{\xi \to 0}\over= \frac{G \; m \; M_{BH}}{T_0 \; l_0 \; \xi} \equiv \frac{\xi_0}{\xi} = \frac{r_0}{r}
 \; \quad , \; \quad \xi_0 \equiv \frac{G \; m \; M_{BH}}{T_0 \; l_0} = \frac{r_0}{l_0} \; .
\ee
That is, the presence of a galactic central black hole implies near the center $ \xi \to 0 $, a $ \xi_0/\xi $
behavior in  $ \nu(\xi) $. $ \xi_0 $ is proportional to the black hole mass $ M_{BH} $.
Recall that in the absence of the black hole 
$ \nu(\xi) $ is bounded for $ \xi \to 0 $ \cite{newas,astro, eqesta}. 

\medskip

We see from eq.(\ref{NUBH}) that in the vicinity $ \xi \lesssim \xi_0 $
of the central black hole,
the chemical potential $ \nu(\xi) $ is dominated by its boundary expression eq.(\ref{NUBH})
and therefore $ \nu(\xi) $ takes positive values  $ \nu(\xi) \gtrsim 1 $. 
Values of $ \nu(\xi) $ larger than unity correspond to a fermionic WDM gas in a
quantum regime \cite{ll}. This is an {\bf important result}: in the vicinity of the central black hole
the fermionic WDM is always in a quantum regime while far from the central black hole
the WDM follows a classical Boltzmann regime \cite{eqesta}. This is natural to understand: 
the strong attractive gravitational force near the central BH compacts the WDM 
and its high density makes it to behave quantum mechanically. On the contrary, far from the BH the gravitational forces are weak,
the WDM is diluted and it is then described by a classical Boltzmann gas.

Ultracompact dwarf galaxies also exhibit WDM in a quantum regime \cite{newas}, \cite{astro}, \cite{eqesta}.

\medskip

$ \nu(\xi) $ takes large positive values for $ \xi \ll \xi_0 $ as implied by eq.(\ref{NUBH}), 
then decreases till vanishing at $ \xi = \xi_A $,  $ \nu(\xi_A) = 0 $  and becomes negative for $ \xi > \xi_A $,
as shown by our detailed resolution of the Thomas-Fermi equation [sec. \ref{3eje} and Fig. \ref{fnu}].

{\vskip 0.1cm} 

Therefore, $ r_A = l_0 \; \xi_A $ plays the role of the {\bf quantum DM radius} of the galaxy
for galaxies exhibiting a central black hole.
Namely, inside $ r_A $ the WDM gas is a {\bf quantum} gas, while
for $ r \gtrsim r_A $ the WDM gas is a classical Boltzmann gas.

\medskip

That is, a small {\bf quantum core} of DM forms around the central black hole.
The size $ r_A $ of the quantum core turns to be smaller for increasing galaxy masses
and black-hole masses because the larger is the black hole mass, the larger is its 
gravitational attraction on the WDM which is thus more compact and hence smaller is the quantum radius core
 $ r_A $.

{\vskip 0.1cm} 

$ r_A $ runs between 0.07 pc to 1.90 pc for galaxies with virial masses
from $ 10^{16} \; M_\odot $ to $ 10^7 \; M_\odot $ [see sec. \ref{3eje}].
In any case, $ r_A $ is much {\bf larger} than the Schwarzchild radius of the central black hole
which runs from $ 10^{-4} $ pc to $ 10^{-8} $ pc.

\bigskip 

In the vicinity of the black hole, the gravitational force due to the black hole
is larger than the gravitational force exerced by the dark matter. 
The influence radius of the black hole $ r_i $  is defined as the radius where both forces
are of equal strength. Notice, that both forces point inwards and always sum up.

\medskip

The total gravitational potential $ V(r) $ and its derivative $ V'(r) $ are given by
\be
V(r) = - \frac{G \; M_{BH}}{r} + \phi(r) \quad , \quad V'(r) = \frac{G \; M_{BH}}{r^2} - \frac1{m} \mu'(r)
\ee
where we used eq.(\ref{potq}). In dimensionless variables $ V'(r) $ becomes
\be
 V'(r) = \frac{T_0}{m \; l_0 \; \xi} \left[\frac{\xi_0}{\xi} - \xi \; \frac{d\nu}{d\xi}\right]
\ee
The black hole and dark matter gravitational forces become equal at $ \xi = \xi_i $.
$ \xi_i $  is the solution of the equation
\be\label{dnudxi}
\xi_i^2 \; \left. \frac{d\nu}{d\xi}\right|_{\xi_i} = \xi_0 \; .
\ee

In the Thomas-Fermi approach to galaxies with a central supermassive black hole,
the boundary condition for $ \nu(\xi) $ at $ \xi \to 0 $ imposes the black hole presence
according to eq.(\ref{NUBH}). That is,
\be\label{conbh}
\nu(\xi) \buildrel_{\xi \to 0}\over= \frac{\xi_0}{\xi} + A + {\cal O}(\xi)
\ee
where $ \xi_0 $ is the dimensionless radius defined in  eq.(\ref{NUBH})
and $ A $ is a constant that determines the properties of the corresponding
galaxy solution as the galaxy mass and galaxy radius. In the absence of the central BH, we have $ \xi_0 = 0 
\; , \; \nu(0) = A $ and the boundary condition used in refs. \cite{newas}, \cite{astro},  \cite{eqesta} is recovered.

\subsection{Main physical magnitudes of the galaxy plus central black hole system}

We find the main physical galaxy magnitudes, such as the
mass density $ \rho(r) $, the velocity dispersion $ \sigma^2(r) = v^2(r)/3 $ and the pressure 
$ P(r) $, which are all $r$-dependent, as: 
\bea\label{gorda}
&& \rho(r) = \frac{m^\frac52}{3 \, \pi^2 \; \hbar^3} \; \left(2 \; T_0\right)^\frac32 \; I_2(\nu(\xi)) =
\rho_0 \; \frac{I_2(\nu(\xi))}{I_2(\nu_0)} 
\;, \;  \rho_0 = \frac{m^\frac52}{3 \, \pi^2 \; \hbar^3} \; \left(2 \; T_0\right)^\frac32 \; I_2(\nu_0) 
 \\ \cr \cr 
&& P(r) = \frac{m^\frac32}{15 \, \pi^2 \; \hbar^3} \; \left(2 \; T_0\right)^\frac52 \; I_4(\nu(\xi))
= \frac1{5} \; \left(9 \; \pi^4\right)^{\!\frac13} \; \left(\frac{\hbar^6}{m^8}\right)^{\! \frac13} \;
\left[\frac{\rho_0}{I_2(\nu_0)}\right]^{5/3}  I_4(\nu(\xi)) \label{pres1}\\ \cr \cr 
&& 
I_4(\nu) \equiv 5 \; \int_0^{\infty} \frac{y^4 \; dy}{\sqrt{1 +\frac{2 \, y^2}{\tau}}} \; 
\Psi_{FD}\left(\displaystyle \tau \left[\sqrt{1 +\frac{2 \, y^2}{\tau}}-1\right] -\nu\right) \; .
\eea

As a consequence, from eqs.(\ref{dmu}), (\ref{varsd}), (\ref{varsd2}), and (\ref{gorda}) 
the total WDM mass $ M(r) $ enclosed in a sphere of radius $ r $ (not including the central black hole mass) turns to be
\be\label{cero} 
 M(r) = 4 \, \pi \; \frac{\rho_0\; l_0^3}{I_2(\nu_0)}\,\int_0^{\xi}
    dx\, x^2 \,I_2(\nu(x)) \; .
\ee
That is, $ M(r) $ is the mass enclosed inside a sphere of radius $ r $
not including the mass of the central black hole mass.
 
{\vskip 0.1cm} 

The integral eq.(\ref{cero}) can be computed in closed form by integrating both sides of eq.(\ref{nu})
\be\label{uno}
M(r)= 4 \, \pi \; \frac{\rho_0 \; l_0^3}{I_2(\nu_0)} \left\{
    \xi^2 \; |\nu'(\xi)| - \left[ \left. \xi^2 \; |\nu'(\xi)| \right|_{\xi \to 0}\right]\right\}
\ee
The contribution here from $ \xi \to 0 $ is obtained from the boundary condition 
eq.(\ref{conbh}) with the result
\bea\label{dos}
&& M(r)=  M_0 \; \xi^2 \; |\nu'(\xi)|
    \; \left(\frac{{\rm keV}}{m}\right)^{\! \! 4} \; 
\sqrt{\frac{\rho_0}{I_2(\nu_0)}\; \frac{{\rm pc}^3}{M_\odot}} - M_{BH} \; , \cr \cr 
&& M_0 = 4 \; \pi \; M_\odot \; \left(\frac{R_0}{\rm pc}\right)^{\! \! 3} 
    = 0.8230 \; 10^5 \; M_\odot \; \label{emeder} \; .
\eea
In absence of the central black hole we recover the expression for the total mass $ M(r) $
obtained in  ref. \cite{eqesta}.

\medskip

In these expressions, we have systematically eliminated the energy scale $ T_0 $ 
in terms of the central density $ \rho_0 $ through eq.(\ref{gorda}). 

\medskip

We define the core size $ r_h $ of the halo by analogy with the Burkert density profile as
\be\label{onequarter}
  \frac{\rho(r_h)}{\rho_0} = \frac14 \quad , \quad  r_h = l_0 \; \xi_h \; .
\ee

\medskip

It must be noticed that the surface density 
 \be\label{densu}
\Sigma_0 \equiv  r_h  \; \rho_0  \; ,
\ee
is found nearly {\bf constant} and independent  of 
luminosity in  different galactic systems (spirals, dwarf irregular and 
spheroidals, ellipticals) 
spanning over $14$ magnitudes in luminosity and  over different 
Hubble types. More precisely, all galaxies seem to have the same value 
for $ \Sigma_0 $, namely $ \Sigma_0 \simeq 120 \; M_\odot /{\rm pc}^2 $
up to $ 10\% - 20\% $ \citep{dona,span,kor}.   
It is remarkable that at the same time 
other important structural quantities as $ r_h , \; \rho_0 $, 
the baryon-fraction and the galaxy mass vary orders of magnitude 
from one galaxy to another.

{\vskip 0.1cm} 

The constancy of $ \Sigma_0 $ seems unlikely to be a mere coincidence and probably
reflects a physical scaling relation between the mass and halo size of galaxies.
It must be stressed that $ \Sigma_0 $ is the only dimensionful quantity
which is practically constant among the different galaxies.

\medskip

It is then useful to take here the dimensionful quantity $ \Sigma_0 $ as physical scale 
to express the galaxy magnitudes in the Thomas-Fermi approach. 
That is, we replace the central density $ \rho_0 $
in the above galaxy magnitudes eqs.(\ref{varsd2})-(\ref{emeder}) in terms of 
$ \Sigma_0 $ eq.(\ref{densu}) with the following results
\be\label{E0}
\begin{split}
&l_0 = \left(\frac{9 \; \pi}{2^9}\right)^{\! \! \frac15} \;
\left(\frac{\hbar^6}{G^3 \; m^8}\right)^{\! \! \frac15} \; 
\; \left[\frac{\xi_h \;  I_2(\nu_0)}{\Sigma_0}\right]^{\! \! \frac15} \;, \\ \\
&l_0 = \; 4.2557 \; \left[\xi_h \; I_2(\nu_0)\right]^{\! \frac15} \;
\left(\frac{2 \, {\rm keV}}{m}\right)^{\! \! \frac85} \; 
\left(\frac{120 \; M_\odot}{\Sigma_0 \;  {\rm pc}^2}\right)^{\! \! \frac15}  \;
{\rm pc}\; 
\end{split}
\ee
\be \label{T0}
\begin{split}
&T_0 = \left(18 \; \pi^6 \; \frac{\hbar^6 \; G^2}{m^3}\right)^{\! \frac15}
\; \left[\frac{\Sigma_0}{\xi_h \; I_2(\nu_0)}\right]^{\! \! \frac45} \;, \\ \\
&T_0 =\;\frac{7.12757\;10^{-3}}{
\left[\xi_h \; I_2(\nu_0)\right]^{\frac45}} \;  \left(\frac{2 \, {\rm keV}}{m}\right)^{\! \! \frac35} \; 
\left(\frac{\Sigma_0 \;  {\rm pc}^2}{120 \; M_\odot}\right)^{\! \! \frac45}  \; {\rm K} \;  .
\end{split}
\ee

The dimensionless quantum radius of the galaxie $ \xi_0 $ eq.(\ref{NUBH}) can be expressed as
\be\label{xi0num}
\xi_0 = \left(\frac{2^8}{3^4 \; \pi^7}\right)^\frac15 \; \left[\frac{\xi_h \; I_2(\nu_0)}{\Sigma_0}\right]^{\! \! \frac35}
\; G^\frac65 \; m^\frac{16}5 \; M_{BH} \; ,\ee
\be
\xi_0 = 36.6145 \; 
\left[\xi_h \; I_2(\nu_0) \; \frac{120 \; M_\odot}{\Sigma_0 \;  {\rm pc}^2} \right]^{\! \! \frac35} \;
 \left(\frac{m}{2 \, {\rm keV}}\right)^{\! \! \frac{16}5} \; \frac{M_{BH}}{10^6 \;  M_\odot} \; .
\ee
Moreover, we can express from here the black hole mass as
\be\label{MBHxi0}
M_{BH} = 2.73116 \; 10^4 \; M_\odot \; \frac{\xi_0}{\left[\xi_h \; I_2(\nu_0)\right]^{\frac35}} \;
\left(\frac{\Sigma_0 \;  {\rm pc}^2}{120 \; M_\odot}\right)^{\! \! \frac35} 
 \; \left(\frac{2 \, {\rm keV}}{m}\right)^{\! \! \frac{16}5} \; .
\ee
Furthermore, we get
\bea\label{rMr}
&&r = 4.2557 \;  \xi \; \left[ \xi_h \; I_2(\nu_0) \right]^{\frac15} \;
 \left(\frac{2 \, {\rm keV}}{m}\right)^{\! \! \frac85}  \; 
\left(\frac{120 \; M_\odot}{\Sigma_0 \;  {\rm pc}^2}\right)^{\! \! \frac15} \; {\rm pc} \;   \\ \cr \cr
&&\rho(r) = \left(\frac{2^9 \; G^3 \; m^8}{9 \; \pi \; \hbar^6}\right)^{\! \! \frac15} \; 
\left[\frac{\Sigma_0}{\xi_h \; I_2(\nu_0)}\right]^{\! \frac65} \; I_2(\nu(\xi)) ,  \\ \cr \cr
&&\rho(r) = 18.1967 \ ; \frac{I_2(\nu(\xi))}{\left[\xi_h \; I_2(\nu_0)\right]^{\! \frac65}}
\; \left(\frac{m}{2 \, {\rm keV}}\right)^{\! \! \frac85} \; 
\left(\frac{\Sigma_0 \;  {\rm pc}^2}{120 \; M_\odot}\right)^{\! \! \frac65} \;
\frac{M_\odot}{{\rm pc}^3} \;  \label{rhor} \; \\ \cr \cr
&&M(r) + M_{BH} = 4 \, \pi \;  
\left(\frac{9 \; \pi \; \hbar^6}{2^9 \; G^3 \; m^8}\right)^{\! \! \frac25} \; 
\left[\frac{\Sigma_0}{\xi_h \; I_2(\nu_0)}\right]^{\! \! \frac35}
\; \xi^2 \; |\nu'(\xi)|  , \\ \cr \cr
&&M(r) + M_{BH} =\frac{27312 \; \xi^2}{\left[ \xi_h \; I_2(\nu_0) \right]^{\frac35}}
\; |\nu'(\xi)| \; \left(\frac{2 \, {\rm keV}}{m}\right)^{\! \! \frac{16}5}
\left(\frac{\Sigma_0 \; {\rm pc}^2}{120 \; M_\odot}\right)^{\! \! \frac35}  
M_\odot \;  \label{emer} \\ \cr \cr
&&\sigma^2(r) = \frac13 \; v^2(r) = 
\frac{11.0402}{\left[\xi_h \; I_2(\nu_0)\right]^{\! \frac45}} \; 
\frac{I_4(\nu(\xi))}{I_2(\nu(\xi))} \; 
\left(\frac{2 \, {\rm keV}}{m}\right)^{\! \! \frac85} \;
\left(\frac{\Sigma_0 \;  {\rm pc}^2}{120 \; M_\odot}\right)^{\! \! \frac45}  \;
\left(\frac{\rm km}{\rm s}\right)^2 \; \; , \label{sigmaTF} \\ \cr \cr
&&P(r) = \frac{8 \, \pi}5 \; G \; \left[\frac{\Sigma_0}{\xi_h \; I_2(\nu_0)}\right]^2
\;  I_4(\nu(\xi)) \;,  \\ \cr \cr
&&P(r) = \frac{200.895}{\left[\xi_h \; I_2(\nu_0)\right]^2} \; I_4(\nu(\xi)) \; 
\left(\frac{\Sigma_0 \;  {\rm pc}^2}{120 \; M_\odot}\right)^2 \; \frac{M_\odot}{{\rm pc}^3} \;
 \left(\frac{\rm km}{\rm s}\right)^2 \; \; .\label{presion}
\eea
That is, $ M(r) + M_{BH} $ is the total mass inside a sphere of radius $ r $
including the mass of the central black hole.

\medskip

Notice that both $ M(r) $ and $ M_{BH} $ at fixed $ \Sigma_0 $ {\it do scale} with the WDM
particle mass as $ m^{-\frac{16}5} $.

\medskip 

In particular, the halo galaxy mass $ M_h $ follows from eq.(\ref{emer}) at $ r = r_h $: 
\be\label{Mrh}
M_h \equiv  M(r_h) + M_{BH} = \frac{27312 \; \xi_h^{\frac75}}{\left[I_2(\nu_0) \right]^{\frac35}}
\; |\nu'(\xi_h)| \; \left(\frac{2 \, {\rm keV}}{m}\right)^{\! \! \frac{16}5}
\left(\frac{\Sigma_0 \; {\rm pc}^2}{120 \; M_\odot}\right)^{\! \! \frac35}  M_\odot \; .
\ee

\medskip

The phase-space density $ Q(r) $ follows from eqs.(\ref{rhor}) and (\ref{sigmaTF}) as
\be\label{RTF}
Q(r) \equiv \frac{\rho(r)}{\sigma^3(r)} = 3 \; \sqrt3 \; \frac{\rho(r)}{<v^2>^\frac32(r)} 
= \frac{\sqrt{125}}{3 \; \pi^2} \; \; \frac{m^4}{\hbar^3}  \;
\frac{I_2^{\frac{5}{2}}(\nu(\xi))}{I_4^{\frac32}(\nu(\xi))}  \; .
\ee
Notice that $ Q(r) $ turns to be independent of $ T_0 $ and $ \Sigma_0 $.
In addition, $ Q(r)/m^4 $ has no explicit dependence on the DM particle mass.

\medskip

For a fixed value of the surface density $ \Sigma_0 $, the 
solutions of the Thomas-Fermi eqs.(\ref{nu}) are parametrized by two
parameters: the dimensionless central radius $ \xi_0 $ and the constant $ A $
characteristic of the chemical potential behaviour eq.(\ref{conbh}) at the center $\xi \to 0 $ .

\medskip

Also, at fixed surface density $ \Sigma_0 $, the halo mass $ M_h $, the black hole mass $ M_{BH} $,
the characteristic lenght $ l_0 $, the density  $ \rho_0 $ and the effective temperature $ T_0 $ are only functions of 
$ \xi_0 $ and the constant $ A $.

\medskip

The circular velocity $ v_c(r) $ is defined through the virial theorem as
\be\label{vci}
 v_c(r) \equiv \sqrt{\frac{G \; \left[M(r) + M_{BH} \right]}{r}} \; ,
\ee
and it is directly related by eq.(\ref{dmu}) to the derivative of the chemical potential as 
\be\label{vcexp}
v_c(r) = \sqrt{- \frac{r}{m} \; \frac{d\mu}{dr}} = \sqrt{-\frac{T_0}{m} \; \frac{d\nu}{d\ln \xi}} \; .
\ee
Expressing $ T_0 $ in terms of the surface density  $ \Sigma_0 $ using
eq.(\ref{T0}) we have for the circular velocity the explicit expression
\be\label{vtf}
v_c(r) = 5.2537 \; \frac{\sqrt{-\xi \; \nu'(\xi)}}{\left[ \xi_h \; I_2(\nu_0) \right]^{\frac25}}
\; \left(\frac{2 \, {\rm keV}}{m}\right)^{\! \! \frac45} \;
\left(\frac{\Sigma_0 \;  {\rm pc}^2}{120 \; M_\odot}\right)^{\! \! \frac25}  \; \frac{\rm km}{\rm s} \; .
\ee

\medskip

For $ r \to 0 $ the circular velocity $ v_c(r) $ grows due to the black hole field as
\be\label{vtf0}
v_c(r) \buildrel_{r \to 0} \over= \sqrt{\frac{T_0}{m} \; \frac{r_0}{r}} \; ,
\ee
where we used eqs.(\ref{fimu0}) and (\ref{vcexp}).

\subsection{Galaxy properties in the diluted Boltzmann regime}

In the diluted Boltzmann regime, $ \nu_0 \lesssim -5 $ corresponding to large galaxies 
 $ M_h \gtrsim 10^6 \; M_\odot $, we find for the main galaxy magnitudes the following analytic expressions:
\bea\label{dilu}
&&M_h = 1.75572 \; \Sigma_0 \; r_h^2 \quad, \quad 
r_h =  68.894 \; \sqrt{\frac{M_h}{10^6 \; M_\odot}
\frac{120\; M_\odot}{\Sigma_0 \;  {\rm pc}^2}} \;\; \; {\rm pc}\;   \\ \cr \cr
&&T_0 = 8.7615 \; 10^{-3} \; \sqrt{\frac{M_h}{10^6 \; M_\odot}} \; \frac{m}{2 \, {\rm keV}}
\; \sqrt{\frac{\Sigma_0 \;  {\rm pc}^2}{120 \; M_\odot}} \; {\rm K} \quad \label{T0gran}
\\ \cr \cr
&&\rho(r) = 5.19505 \; \left(\frac{M_h}{10^4 \; M_\odot} \; \frac{\Sigma_0 \;  {\rm pc}^2}{120 \; M_\odot}
\right)^{\! \! \frac34} \; \left(\frac{m}{2 \, {\rm keV}}\right)^4 \; e^{\nu(\xi)} \; \; 
\frac{M_\odot}{{\rm pc}^3} \; \\ \cr \cr
&&v_c^2(r) = 33.9297 \; \sqrt{\frac{M_h}{10^6 \; M_\odot} \;
\frac{\Sigma_0 \;  {\rm pc}^2}{120 \; M_\odot}} \;
\left| \frac{d \nu(\xi)}{d \ln \xi}\right| 
\; \left(\frac{\rm km}{\rm s}\right)^2 \; , \; \\ \cr \cr
&&v_c^2(r_h) = 62.4292 \; \sqrt{\frac{M_h}{10^6 \; M_\odot} \;
\frac{\Sigma_0 \;  {\rm pc}^2}{120 \; M_\odot}} \;
\; \left(\frac{\rm km}{\rm s}\right)^2 \; \label{vch} \\ \cr \cr
&&M(r) + M_{BH} = 7.88895 \; \left| \frac{d \nu(\xi)}{d \ln \xi}\right|  \; \frac{r}{\rm pc}
\; \sqrt{\frac{M_h}{10^6 \; M_\odot} \; \frac{\Sigma_0 \;  {\rm pc}^2}{120 \; M_\odot}} \; .
\eea
Eqs.(\ref{T0}) and (\ref{T0gran}) allow us to express the quantity $ \xi_h \; I_2(\nu_0) $ in terms of the observable galaxy
magnitudes $ M_h $ and $ \Sigma_0 $ for large galaxies $ M_h \gtrsim 10^6 \; M_\odot $ in the diluted regime.
We obtain from eqs.(\ref{T0}) and (\ref{T0gran})
\be\label{xihI}
\xi_h \; I_2(\nu_0) = 0.772598 \; \left(\frac{10^6 \; M_\odot}{M_h}\right)^{\! \! \frac58} \;
\left(\frac{2 \, {\rm keV}}{m}\right)^2 \; \left(\frac{\Sigma_0 \; {\rm pc}^2}{120 \; M_\odot}\right)^{\! \! \frac38}
\ee

It is illuminating to express the radius $ r_0 $ eq.(\ref{NUBH}) in terms
of $ M_{BH} $ and $ \Sigma_0 $ for large galaxies $ M_h \gtrsim 10^6 \; M_\odot $.
It follows from eqs.(\ref{E0}), (\ref{xi0num}) and (\ref{xihI}) that,
\be
r_0 = l_0 \; \xi_0 = 126.762 \; \sqrt{\frac{10^6 \; M_\odot}{M_h}} \; \frac{M_{BH}}{10^6 \; M_\odot} \; 
\sqrt{\frac{120 \; M_\odot}{\Sigma_0 \; {\rm pc}^2}} \; \; {\rm pc} 
\ee

This explicitely provides the value of the radius $ r_0 $ in terms of the black hole mass $M_{BH}$, the halo mass $ M_h $, and the reference surface density $\Sigma_0$.

\medskip

In summary, we see the power of the WDM Thomas-Fermi approach to describe
the structure and the physical state of galaxies
in a clear way and in very good agreement with observations.

\section{Explicit Thomas-Fermi Galaxy solutions with central supermassive black holes}

We solve here the Thomas-Fermi equations (\ref{nu}) with the boundary conditions (\ref{conbh})
for a galaxy with a central black hole.

\subsection{Local thermal equilibrium in the Galaxy}\label{locteq}

In ref. \cite{eddi} using the Eddington equation for dark matter in galaxies and observed
density profiles, it is shown that the DM halo is realistically a self-gravitating thermal gas for
$ r \lesssim R_{virial} $. More precisely, the DM halo can be consistently considered in a local thermal 
equilibrium situation with: {\bf(i)} a constant temperature $ T_0$ for $ r \lesssim 3 \; r_h $, and  {\bf(ii)} a space dependent temperature $ T(r) $ for $ 3 \; r_h < r \lesssim R_{virial} $, which slowly decreases with $ r $.
$ T(r) $ outside the halo radius nicely follows the decrease of the circular velocity squared 
$ T_c(r) $ \cite{eddi}. These results are  physically understood because thermalization is more easy achieved in the inner 
regions due to the fact that the gravitational interaction
is stronger than in the external regions where  instead
virialization occurs. The slow decreasing of the
temperature $T(r)$ with the halo radius consistently
corresponds to a transfert flux of the kinetic energy into
potential energy. These results were derived from empirical
observed density profiles which do not
have information of the regions near the central
black hole.

{\vskip 0.2cm}

The constant temperature $ T_0 $ for $ r \lesssim 3 \; r_h $ turns to be in the Kelvin
scale for a DM particle mass in the keV scale \cite{eqesta}.

\medskip

To implement the Thomas-Fermi approach for a galaxy plus a central black hole
we take into account the results of Ref. \cite{eddi}.
We simply set the WDM temperature to be a constant $ T_0 $ except in the vicinity of the central black hole.
We do not assume WDM thermalization near the central black hole where the  black hole force is strong but 
we assume virialization. Namely, the WDM square velocity is determined by the  black hole gravitational field
through virialization.

{\vskip 0.2cm}

In summary,
\begin{itemize}

\item {Near the central black hole,  the space dependent 
temperature is given by equipartition and the virial theorem
\be\label{tcbh}
T_c(r) = \frac{m}3 \; v_c^2(r) = \frac{G \; m}{3 \; r} \; M_{BH} = \frac{T_0 \; \xi_0}{3 \; \xi}
\ee
where we used eqs.(\ref{varsd}) and (\ref{NUBH}). We use this temperature 
$ T_c(r) $ for $ \xi \leq \xi_0/3 $. $ T_0 $ is given by eq.(\ref{T0}).}

\item {For $  \xi \geq  \xi_0/3  $ we set
$$
T_c(r) = T_0 \; .
$$}

Here the circular temperature $ T_c(r) $ associated to the velocity squared is given by
\be\label{tclong}
T_c(r) = \frac{m}3 \; \frac{G \; [M(r)+M_{BH}] }{r} \; ,
\ee
where $ M(r) + M_{BH} $, the mass of the galaxy inside the radius $ r $ including the BH mass $ M_{BH} $ ,
is given by eq.(\ref{emer}). Inserting eq.(\ref{emer}) into eq.(\ref{tclong}) and using eq.(\ref{T0}) yields,
\be\label{tclo2}
 T_c(r) = \frac13 \; \xi \; |\nu'(\xi)| \; T_0 \; .
\ee
Near the central black hole, that is for $ \xi \leq \xi_0/3 $, the chemical potential $ \nu(\xi) $ is given by eq.(\ref{NUBH}).
Inserting eq.(\ref{NUBH}) into eq.(\ref{tclo2}) yields eq.(\ref{tcbh}) as it must be.

{\vskip 0.2cm}

\item {We find from our extensive numerical calculations that the halo is thermalized at the uniform
temperature $ T_0 $ and matches the circular temperature  $ T_c(r) $ by $ r \sim 3 \; r_h $.
This picture is similar to the picture found in the absence of the central black hole which follows from the observed density profiles in the Eddington-like approach to galaxies \cite{eddi}.
We obtain here in the Thomas-Fermi approach and in the presence of a central supermassive black hole that:
the halo is thermalized at a uniform temperature $ T_0 $ inside $ r \lesssim 3 \; r_h $ which matches
the circular temperature  $ T_c(r) $ at $ r \sim 3 \; r_h $ [see Fig.\ref{vel}].}

{\vskip 0.2cm}

\item {In summary, each galaxy solution with a central black hole depends only on {\bf two
free parameters}: the dimensionless constants $ \xi_0 $ and $ A $ in eq.(\ref{conbh}).
We have a two-parameter family of Thomas-Fermi galaxy solutions with a central supermassive black hole
parametrized by $ \xi_0 $ and $ A $.}

{\vskip 0.2cm}

The black hole mass $ M_{BH} $ {\it grows} when $ \xi_0 $ grows as shown by eq.(\ref{MBHxi0}).
Notice that $ M_{BH} $ does not simply grow linearly with $ \xi_0 $ due to the presence 
of the factor
$$
\left[\xi_h \; I_2(\nu_0)\right]^{-\frac35},
$$
in eq.(\ref{MBHxi0}). 

\end{itemize}

{\vskip 0.2cm}

From our extensive numerical calculations we find that the galaxy mass increases and the galaxy size
increases when the constant $ | A | $  characteristic of the 
the central behaviour of $ \nu(\xi)$ for $ \xi \to 0 $ eq.(\ref{conbh}) increases.
This is similar to the case in the absence of 
central black holes where $ A = \nu(0) $ \cite{newas}, \cite{astro},  \cite{eqesta}.

\subsection{Thomas-Fermi equations with r-dependent temperature  T{c}(r) }\label{tder}

For a $r$-dependent temperature $ T_{c}(r) $ the normalized energy density $ I_2(\nu) $
[recall eq.(\ref{dfI})] takes the form
\be\label{i2tder}
I_2(\nu) = 3 \; \int_0^{\infty} y^2 \; dy \; \sqrt{1 +\frac{2 \, y^2}{\tau}} \; 
\Psi_{FD}\left(\displaystyle \frac{T_0}{T_c(r)}\left[
\tau \left(\sqrt{1 +\frac{2 \, y^2}{\tau}}-1\right) - \nu\right]\right) \quad  , 
\ee
which becomes eq.(\ref{dfI}) for a constant temperature $ T_c(r) = T_0 $, that is in the region
$ \xi \geq \xi_0/3 $, ie $ r \geq r_0/3 $.

{\vskip 0.2cm}

Near the BH, for $ \xi \leq \xi_0/3 $, we have from eq.(\ref{tcbh}),

\be
\frac{T_0}{T_c(r)} = \frac{3 \; \xi}{\xi_0}  \; .
\ee

Beyond $ r = 3 \; r_h $, using eq.(\ref{tclo2}) these quantities take the values

\be
\frac{T_0}{T_c(r)} = \frac{3}{\xi \; |\nu'(\xi)|}
\ee

Notice that  $ T_c(r) $ grows  for $ r \to 0 $  as $ 1/r $ due to the BH presence
eq.(\ref{tcbh}). On the contrary,  $ T_c(r) $ decreases for increasing $ r > 3 \; r_h $.

For a density profile scaling at large $r$, $ r > 3 \; r_h $,  as $ r^{-2\alpha} $  we find
 $ T_c(r) \sim r^{2(1-\alpha)} $. Because observations favour $ \alpha \sim 1.5 > 1 $, 
$ T_c(r) $ decreases for increasing $ r  > 3 \; r_h $, as in the case where the black hole is absent \cite{eddi}.

\subsection{Examples of Thomas-Fermi Galaxy solutions with a central supermassive black hole}\label{3eje}

We consider here three realistic examples: a small mass galaxy, a medium mass and a large mass galaxy.
We choose as boundary conditions in eq.(\ref{conbh}) 
\bea\label{tresgal}
&& \xi_0 = 1 \quad , \quad A = 0 \quad , \quad {\rm small ~ size ~  galaxy} \cr \cr
&& \xi_0 = 7 \quad , \quad A = -10 \quad , \quad {\rm medium ~  size ~  galaxy} \cr \cr
&& \xi_0 = 9 \quad , \quad A = -15\quad , \quad {\rm large ~  size ~  galaxy} 
\; .
\eea
These values are illustrative and yield realistic galaxies with a supermassive central black hole
as we see below. \\
Indeed, we find realistic solutions for a large manifold of boundary conditions.

{\vskip 0.2cm}

To compute these solutions we set as reference values $ m = 2 $ keV, $ \Sigma_0 = 120 \; M_\odot /{\rm pc}^2 $.

{\vskip 0.2cm}

The present solutions allow to characterize the WDM properties that show up in the different
halo regions according to the distance to the central black hole.

{\vskip 0.2cm}

\begin{figure}
\begin{turn}{-90}
\psfrag {"Z10fichi.dat"} {Small Galaxy}
\psfrag{"Z10fimed.dat"}{Medium Galaxy}
\psfrag{"Z10figran.dat"}{Large Galaxy}
\psfrag{Log nu}{$ \log_{10} |\nu(\xi)| $}
\psfrag{log_{10} ri/rh}{$ \log_{10} r/r_h $}
\psfrag{ACA xa}{$ x_A \downarrow $}
\includegraphics[height=12.cm,width=10.cm]{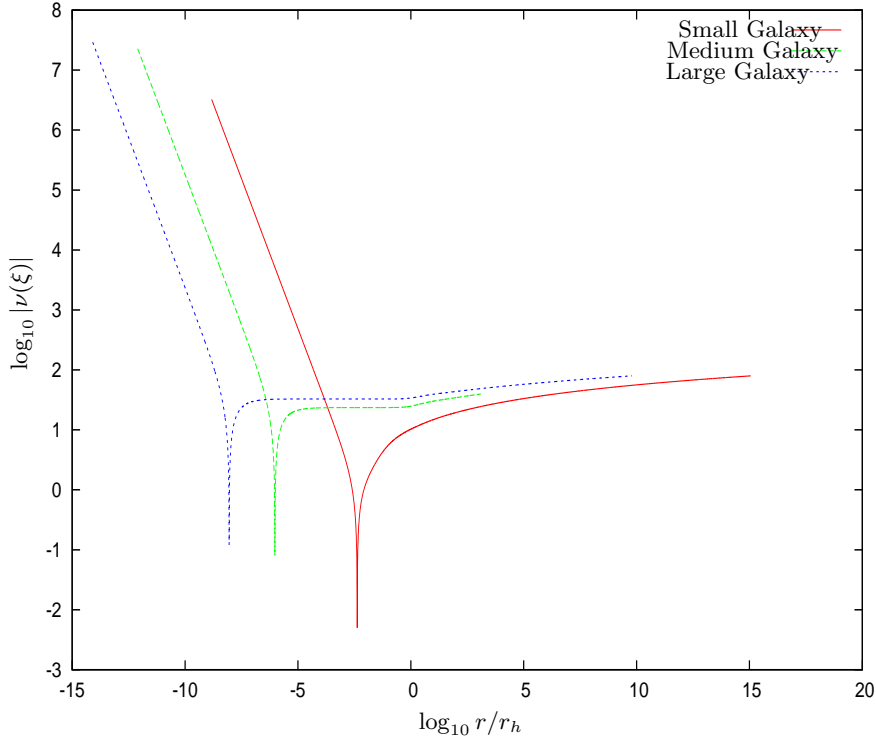}
\end{turn}
\caption{The dimensionless chemical potential $ \log_{10} |\nu(\xi)| $ vs. $ \log_{10}(\xi/\xi_h) = \log_{10} (r/r_h) $
for the three illustrative galaxy solutions with central SMBH defined by eq.(\ref{tresgal}).
$ \nu(\xi) $ is negative for $ r > r_A = l_0 \; \xi_A $ and WDM exhibits there
a classical dilute Boltzmann gas behaviour, while WDM exhibits a compact quantum gas
behaviour for $ r < r_A $ where the chemical potential is positive.
The point $ r_A $ where the chemical potential vanishes {\bf defines the transition
from the quantum to the classical galaxy WDM behaviour}. $ r_A $ is at the downward spike of $ \log_{10} |\nu(\xi)| $
where $ \nu(\xi) $ vanishes.}
\label{fnu}
\end{figure}

\begin{figure}
\begin{turn}{-90}
\psfrag{d nu/dx}{$ \log_{10} [|d\nu(\xi)/dx|/3] $}
\psfrag{"Z10fipchi.dat"}{Small Galaxy}
\psfrag{"Z10fipmed.dat"}{Medium Galaxy}
\psfrag{"Z10fipgran.dat"}{Large Galaxy}
\psfrag{log_{10} ri/rh}{$ \log_{10} r/r_h $}
\psfrag{ACA xi}{$ x_i \downarrow $}
\includegraphics[height=12.cm,width=10.cm]{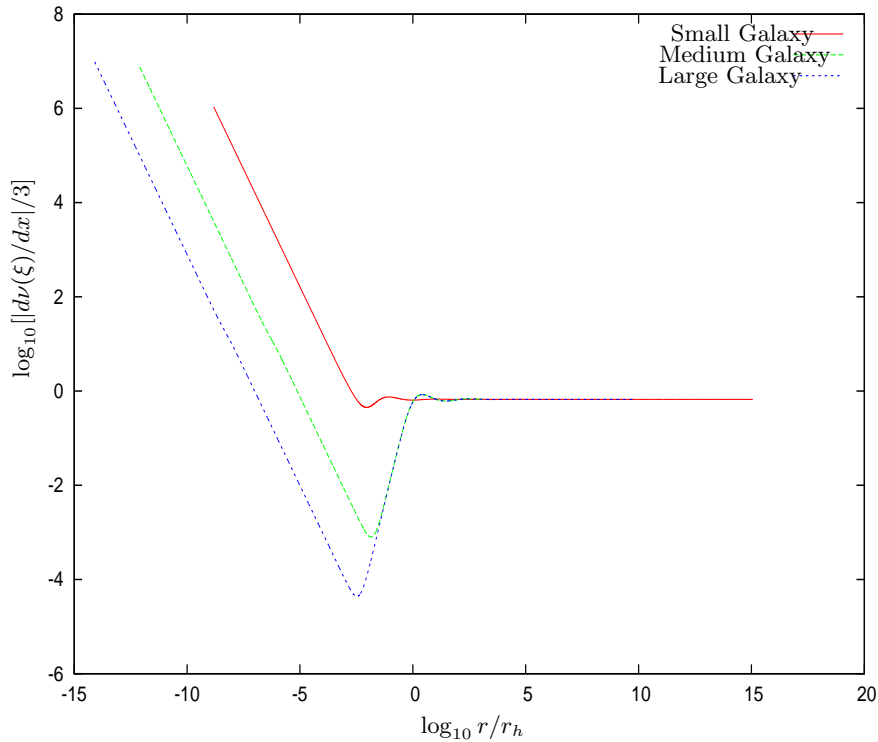}
\end{turn}
\caption{The derivative of the dimensionless chemical potential 
$ \log_{10}  [|d\nu(\xi)/dx|/3] $ vs. $ \log_{10}(\xi/\xi_h) = \log_{10} (r/r_h) $
for the three galaxy solutions with central SMBHs defined by eq.(\ref{tresgal}).
For $ r \lesssim r_i $ both $ \nu(\xi) $ and $ |d\nu(\xi)/dx| $
follow the behaviour dictated by the central black hole. $ r_i $ is the influence radius of the BH defined by eq. (\ref {dnudxi}).
For $ r \gtrsim r_i , \; \nu(\xi) $ and $ |d\nu(\xi)/dx| $ are dominated by WDM
and exhibit a similar behaviour to that for the Thomas-Fermi galaxy solutions without a central
black hole \cite{newas,astro,urc,eqesta}. }
\label{nup}
\end{figure}

\begin{figure}
\begin{turn}{-90}
\psfrag{Log rho/rho0}{$ \log_{10} \rho(r)/\rho_0 $}
\psfrag{"Z10rhochi.dat"}{Small Galaxy}
\psfrag{"Z10rhomed.dat"}{Medium Galaxy}
\psfrag{"Z10rhogran.dat"}{Large Galaxy}
\psfrag{log_{10} ri/rh}{$ \log_{10} r/r_h $}
\psfrag{ACA xa}{$ x_A \downarrow $}
\psfrag{ACA xi}{$ x_i \downarrow $}
\psfrag{ACA xh}{$ x_h \downarrow $}
\includegraphics[height=12.cm,width=10.cm]{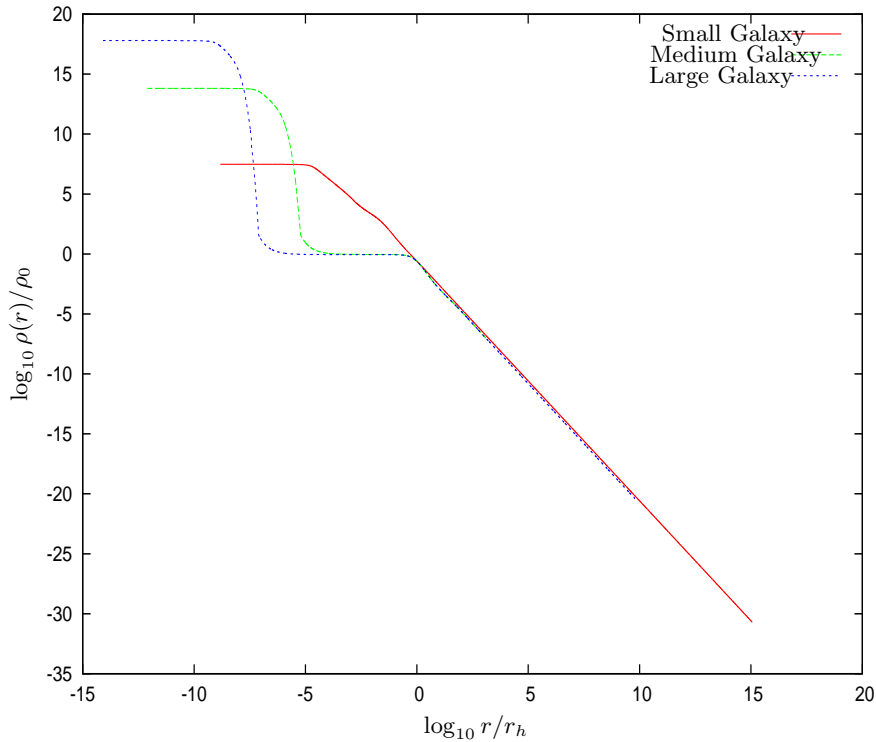}
\end{turn}
\caption{The density $ \rho $ normalized at the influence radius $ r_i $,
$ \log_{10} ( \rho(r)/\rho_0 ) $ vs. $ \log_{10} (r/r_h) $ for the three galaxy solutions with central SMBHs. Notice that in the quantum  gas WDM region $ r < r_A $
the density is constant clearly exhibiting a plateau behaviour corresponding to the quantum Fermi gas behaviour in such region.}
\label{rho}
\end{figure}

For the three representative galaxy solutions eq.(\ref{tresgal}), we plot in Fig. \ref{fnu} the dimensionless chemical potential
$ \; \log_{10} \nu(\xi)$ versus the dimensionless radius $ \log_{10} (\xi/\xi_h) = \log_{10} (r/r_h) $,  $r_h$ being the halo radius (and $\xi_h$ the dimensionless one) ;
in Fig. \ref{nup} we plot the derivative  $ \; \log_{10} |d\nu(\xi)/dx| $ vs.  $ \log_{10} (r/r_h) $,
and in Fig. \ref{rho} we plot  \; the density profiles  $\log_{10} [\rho(\xi)/\rho_0]$ vs. $ \log_{10} (r/r_h) $,
[recall that $ \rho_0 \equiv \rho(\xi_i)$, $\xi_i$ being the dimensionless influence radius of the black hole eq.(\ref{dnudxi}), that is, when the black hole and dark matter gravitational forces become equal].

{\vskip 0.2cm}

Notice that the curves for the three galaxy solutions are of similar size
thanks to the use of the rescaled variable $ r/r_h = \xi/\xi_h $ in the abscissa.
The dimensionless halo radius $ \xi_h $ increases by five orders of magnitude going from the small to the large size galaxy.

{\vskip 0.2cm}

For the relevant parameters of the solutions we obtain the following results:
\bea\label{res3gal}
&& {\rm \bf {Small ~ size ~ galaxy:}} \cr \cr
&& r_i = 221 \; {\rm pc}  \quad , \quad r_h  =  452  \; {\rm pc} \quad, \quad T_0 = 0.0978 \; {\rm K},  \cr \cr   
&& \sqrt{<v^2>}(r \gtrsim  r_A) = 35.48 \; {\rm km/s}, \quad \quad 
\sqrt{<v^2>}(r \lesssim  r_A) = 383.75 \; {\rm km/s} \; , \cr \cr
&& M_h = 7.678 \; 10^7 \; M_\odot, \quad M_{vir} =  \; 8.582 \; 10^{8} \; M_\odot,  \cr \cr
&&M_{BH} =  1.947 \; 10^5 \; M_\odot, \quad r^{Schw}_{BH} = 1.863 \; 10^{-8} \; {\rm pc} \; , \cr \cr
&& \rho_0 = 1.797 \; 10^{-23} \; {\rm g/cm^3}, \cr \cr
&& \rho_{A} = 0.9878 \; 10^{-19}  \; {\rm g/cm^3}, \quad M_{A} = 8.767 \; 10^{4} \; M_\odot, \quad r_A = 1.91  \; {\rm pc}  \;  . \\ \cr \cr 
&& {\rm \bf {Medium ~ size ~ galaxy:}} \cr \cr
&& r_i = 54.3 \; {\rm pc}  \quad , \quad r_h  =  210  \; {\rm kpc} \quad , \quad
T_0 = 26.97 \; {\rm K}, \cr \cr  
&&\sqrt{<v^2>}(r \gtrsim  r_A) = 559.8 \; {\rm  km/s}, \quad \quad
\sqrt{<v^2>}(r \lesssim  r_A) = 6370.9 \; {\rm km/s} \; , \cr \cr
&& M_h = 9.022 \; 10^{12} \; M_\odot, \quad \quad M_{vir} =  \; 8.222 \; 10^{13} \; M_\odot, \cr \cr
&&M_{BH} = 9.224 \; 10^7 \; M_\odot, \quad \quad r^{Schw}_{BH} = 8.828 \; 10^{-6} \; {\rm pc} \label{2res3gal}
\; , \cr \cr 
&& \rho_0 = 3.867 \; 10^{-26} \; {\rm g/cm^3}, \cr \cr
&&\rho_{A} = 7.182 \; 10^{-15} \; {\rm g/cm^3},
\quad M_{A} = 1.932 \; 10^{7}  \; M_\odot, \quad r_A =  0.2 \; {\rm pc} 
\;  . \\ \cr \cr
&& {\rm \bf{Large ~ size ~ galaxy:}} \cr \cr
&& r_i = 21.66 \; {\rm pc}, \quad r_h  =  8.237  \; 10^3 \; {\rm kpc} \quad ,\quad
T_0 = 1061 \; {\rm K}, \cr \cr 
&&\sqrt{<v^2>}(r \gtrsim  r_A) = 3511.2\; {\rm  km/s}, \quad \quad
\sqrt{<v^2>}(r \lesssim  r_A) = 39591\; {\rm km/s}  \; , \cr \cr
&& M_h = 1.3753 \; 10^{16} \; M_\odot, \quad \quad M_{vir} =  \; 3.3482 \; 10^{16} \; M_\odot, \cr \cr 
&&M_{BH} = 1.8632 \; 10^9 \; M_\odot, \quad \quad r^{Schw}_{BH} = 1.783 \; 10^{-4} \; {\rm pc}  \; , \cr \cr
&& \rho_0 = 0.9860 \; 10^{-27} \; {\rm g/cm^3}, \cr \cr
&& \rho_{A} = 2.9163 \; 10^{-12} \; {\rm g/cm^3}, \quad M_{A} = 3.873\; 10^{8} \; M_\odot, \quad r_A = 0.074 \; {\rm pc}\; .\label{grangal}
\eea
$M_A$ stands for the mass inside the radius $r_A$.

\medskip

Notice that the obtained galaxy solutions have halo masses $ M_h > 10^6 \; M_\odot $ and 
therefore belong to the dilute Boltzmann regime \cite{eqesta}. 

\medskip

Let us now analyze the Figures \ref{fnu}-\ref{rho}. We start from the galaxy center and go towards the halo tail.

\begin{figure}
\begin{turn}{-90}
\psfrag{v^2 km/s}{$ \log_{10} <v^2>(r) $ km/s}
\psfrag{"Zv2chi.dat"}{Small Galaxy}
\psfrag{"Zv2med.dat"}{Medium Galaxy}
\psfrag{"Zv2gran.dat"}{Large Galaxy}
\psfrag{"Zv2circhi.dat"}{Circular Velocity Small Galaxy}
\psfrag{"Zv2cirmed.dat"}{Circular Velocity Medium Galaxy}
\psfrag{"Zv2cirgran.dat"}{Circular Velocity Large Galaxy}
\psfrag{"trucho.dat"}{}
\psfrag{log_{10} ri/rh}{$ \log_{10} r/r_h $}
\includegraphics[height=12.cm,width=10.cm]{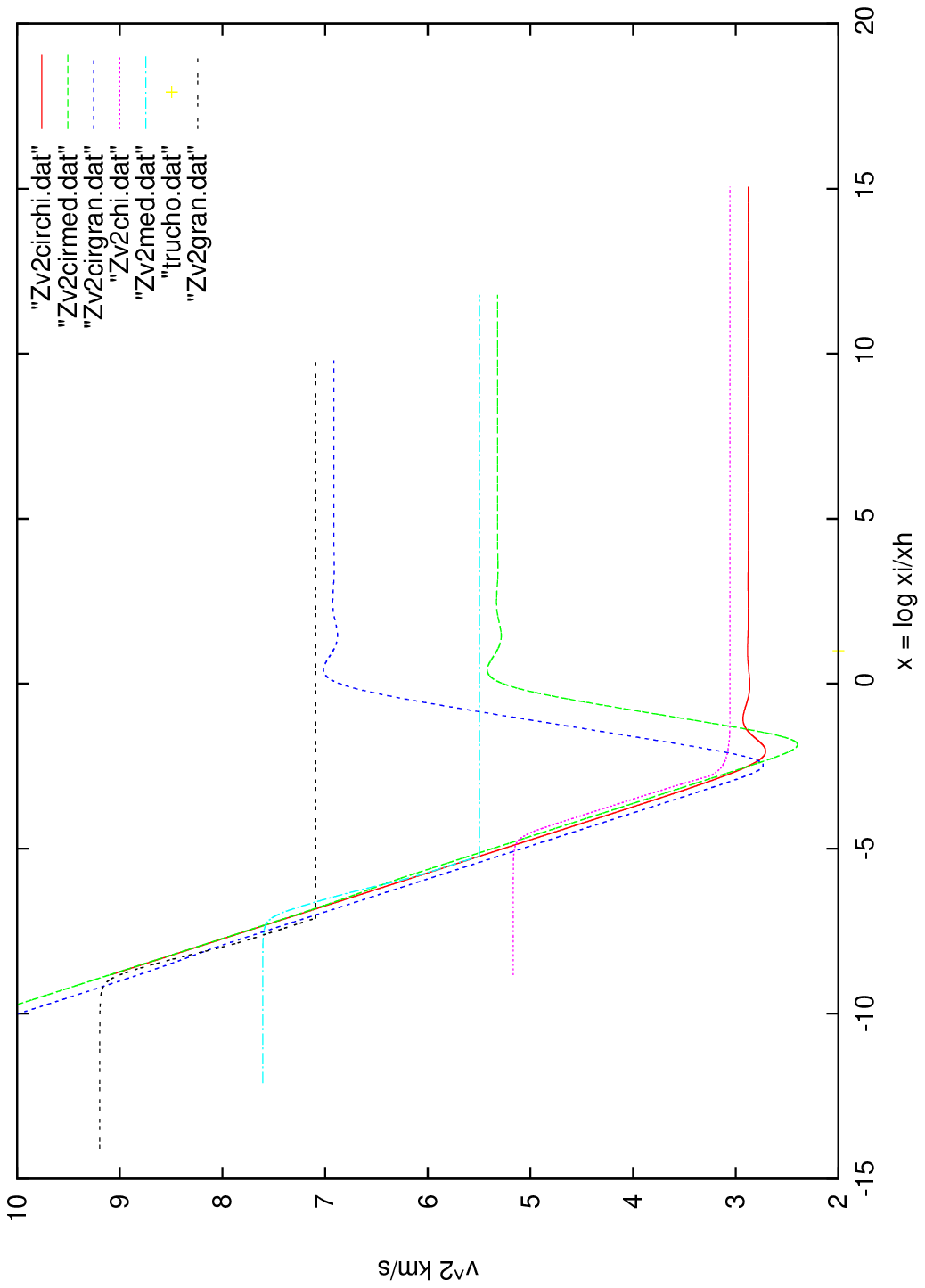}
\end{turn}
\caption{The velocity dispersion $ <v^2>(r) $  and the circular velocity $ v_c^2(r) $ 
for the three representative galaxy solutions with central SMBH vs.  $ \log_{10} (r/r_h) $. 
The velocity dispersion is constant in the Boltzmann and in the quantum regions 
indicating a thermalized WDM with two different temperatures
$ T_ 0 = \frac13\; m \; <v^2>(r) $. For $ r> r_h $ the circular velocity tends to the 
velocity dispersion \cite{eddi}. These results are in agreement with the DM thermalization found in the absence of a central BH \cite{eddi}, \cite{eqesta}.} 
\label{vel}
\end{figure}

\begin{itemize}

\item { {\bf Quantum to classical behaviour}: The central black hole strongly attracts the WDM and 
makes its density very high for $ r < r_A $ where a compact quantum core gets formed.
The dimensionless chemical potential $ \nu(\xi) $ vanishes at $ r = r_A $
and becomes negative for $ r > r_A $. The density $ \rho(r) $ drops several orders of
magnitude immediately after $ r_A $ as shown in Fig. \ref{rho}.
$ \nu(\xi) $ is negative for $ r > r_A $ and the WDM exhibits there
a classical Boltzmann behaviour while the WDM exhibits a quantum 
behaviour for $ r < r_A $ where the chemical potential is large and positive.
Therefore, the point $ r_A $ where the chemical potential vanishes {\bf defines the transition
from the quantum to classical behaviour}. In the quantum region $ r < r_A $ the density exhibits a constant plateau
as shown in Fig. \ref{rho}. Notice from eqs.(\ref{res3gal}) that $ r_A $ turns to be much larger 
than the Schwarzchild radius of the central black hole $  r_A \gg r^{Schw}_{BH} $.}

\item { {\bf Black hole influence radius $ r_i $}: For $ r < r_i $ the black hole gravitational field
dominates over the dark matter gravitational field. The influence radius $ r_i = l_0 \; \xi_i $
is defined by eq.(\ref{dnudxi}). The black hole influence radius turns to be larger than the radius $ r_A $ where
 the chemical potential vanishes, $ r_i > r_A $.
The region $ r_A < r < r_i $ is dominated by the central black hole and
the WDM exhibits there a classical behaviour.
For $ r \lesssim r_i $, we see from Figs. \ref{fnu}-\ref{nup} that
both $ \nu(\xi) $ and $ |d\nu(\xi)/dx| $ follow the behaviour dictated by the central black hole.
That is, from eq.(\ref{conbh})
$$
\nu(\xi) \simeq \xi_0 \; e^{-x} + A = \xi_0 \; \frac{r_h}{r}  + A 
\quad , \quad |d\nu(\xi)/dx| \simeq \xi_0
\; e^{-x} = \xi_0 \; \frac{r_h}{r} \quad ,
\quad x \equiv \ln \frac{r}{r_h} \; ,
$$
which produce straight lines on the left
part of the logarithmic plots Figs.
\ref{fnu}-\ref{nup}.
For $ r \gtrsim r_i , \; \nu(\xi) $ and $
|d\nu(\xi)/dx| $ are dominated by the WDM
and exhibit a similar behaviour to that of
the Thomas-Fermi solutions without a central
black hole \cite{newas}, \cite{astro}, \cite{urc}, \cite{eqesta}. \\
 
Fig. \ref{rho} shows that the local density behaviour is dominated by the black hole
for $ r \lesssim r_i $.  Coherently, for $ r \gtrsim r_i $ the WDM gravitational field
dominates over the black hole field and the galaxy core shows up for $ r_i \lesssim r \lesssim r_h $
in Fig. \ref{rho}. For medium and large galaxies the core is seen as a plateau.
At the same time the chemical potential is negative
for $ r \gtrsim r_i > r_A $ and the WDM is a classical Boltzmann gas in this region.}

\item { {\bf Halo radius $ r_h $}:
Finally, we see in Fig. \ref{rho} the tail of the WDM density profile for $ r \gtrsim r_h $ which
exhibit a similar shape for all three galaxy solutions.}

\item { {\bf WDM thermalization}: As shown by Fig. \ref{vel}, the velocity dispersion $ <v^2>(r) $
is constant as a function of $ r $ indicating a thermalized WDM with temperature
$$
T_ 0 = \frac13\; m \; <v^2> \; .
$$
WDM gets thermalized as in the absence of the central black hole \cite{eqesta}.
This is consistent with the use of  a thermal Fermi-Dirac distribution function for
$ r\geq r_0/3 $.}

\item {We also plot in Fig. \ref{vel} the circular velocity given by eq.(\ref{vtf}) 
vs. $ \log_{10} r/r_h $. For $ r> r_h $ the circular velocity tends to the 
velocity dispersion as obtained from the Eddington equation for realistic
density profiles \cite{eddi}. For $ r \to 0 $ the circular velocity
grows as in eq.(\ref{vtf0}) due to the central black hole field.}

\item {WDM inside a small core of radius $ r_A $ is in a quantum gas high density state, namely a Fermi nearly degenerate state with  nearly constant
density  $ \rho_A $. For the three galaxy solutions, the values of $ r_A $ and $ \rho_A $
are given by eqs.(\ref{res3gal})-(\ref{grangal}). Notice, that the density $ \rho_A $
is orders of magnitude larger than its values for $ r > r_A $ where the WDM is in the classical Boltzmann regime.}

\item {We also give in eqs.(\ref{res3gal})-(\ref{grangal}) the WDM mass $ M_A $ inside  $ r_A $.
$ M_A $ represents only a small fraction of the halo or virial mass of the galaxy
but it is a significant fraction of the black hole mass $ M_{BH} $. 
We see from eqs.(\ref{res3gal})-(\ref{grangal}) that $ M_A $ amounts
for 20\% of  $ M_{BH} $ for the medium and large galaxies and 45\% for the small galaxy.}

\end{itemize}

\bigskip

\subsection{Quantum physics in galaxies}

In order to determine whether a physical system has a classical or quantum nature
one has to compare the average distance between particles $ d $ with their
de Broglie wavelength $ \lambda_{dB} $.

\medskip

The de Broglie wavelength of DM particles in a galaxy can be expressed as
\be\label{LdB}
\lambda_{dB}(r)   = \frac{h}{m \; v(r) } \; ,
\ee
where $ h $ stands for the Planck constant and 
$ v \equiv \sqrt{<v^2>}$  is the velocity dispersion, while the average interparticle distance $ d $ at $ r $ can be estimated as
\be\label{dis}
d (r) = \left( \frac{m}{\rho(r)} \right)^{\! \! \frac13} \; 
\ee
Here $ \rho(r) $ is the local density in the  galaxy core.

\medskip

We can measure the classical or quantum character of the system by considering the ratio
$$ 
{\cal R}(r)  \equiv \frac{\lambda_{dB}(r)}{d(r) }
$$
For $ {\cal R} \lesssim 1 $ the system is of classical dilute nature while for $ {\cal R} \gtrsim 1 $
it is a macroscopic quantum system.

\medskip

By using the phase-space density eq.(\ref{RTF}),
$$
 Q(r)  = \frac{\rho(r)}{\sigma^3(r)} ,
$$
and eqs.(\ref{LdB})-(\ref{dis}), $ {\cal R}(r)  $ can be expressed solely in terms of the phase space-density $ Q(r) $
as \cite{newas}, \cite{astro},\cite{eqesta}
\be\label{defR}
{\cal R}(r)  = \frac{2 \; \pi}{\sqrt3} \; \hbar \; \left( \frac{Q(r)}{m^4}\right)^{\! \! \frac13} \; .
\ee

{\vskip 0.2cm}

\begin{figure}
\begin{turn}{-90}
\psfrag{"ZRchi.dat"}{Small Galaxy}
\psfrag{"ZRmed.dat"}{Medium Galaxy}
\psfrag{"ZRgran.dat"}{Large Galaxy}
\psfrag{R}{$ {\cal R}(r) $}
\psfrag{log r/rh}{$ \log_{10} r/r_h $}
\includegraphics[height=12.cm,width=10.cm]{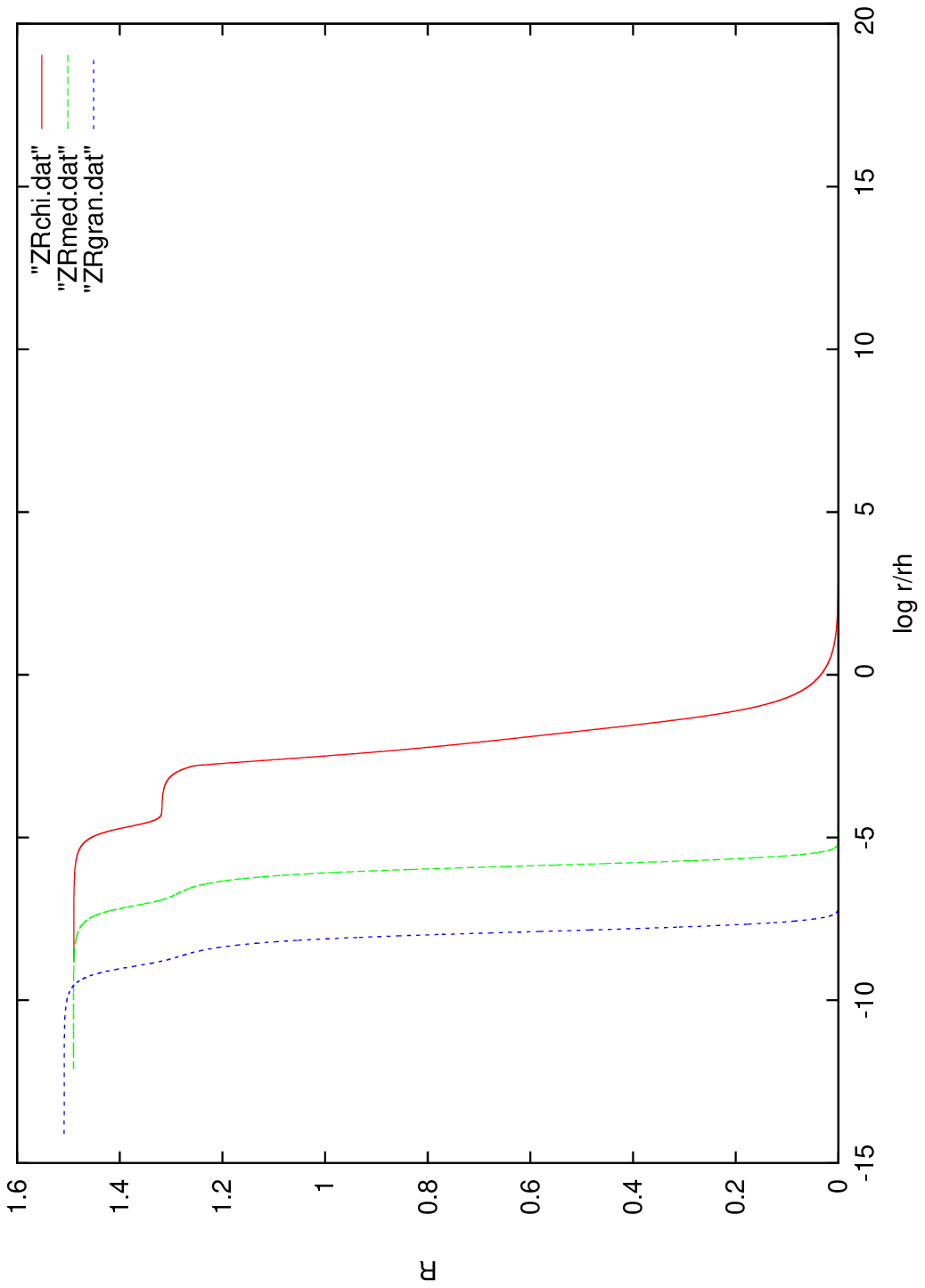}
\end{turn}
\caption{The ratio $ \cal R $ of the particle de Broglie wavelength to the interparticle distance in the galaxy
as a function of $ r $  for the three representative galaxy solutions with central SMBH: Small galaxy (red), Medium galaxy (green), Large galaxy (blue) \\ 
For $ {\cal R} \lesssim 1 $ the galaxy plus SMBH system is of classical nature while for $ {\cal R} \gtrsim 1 $
the system is quantum. The transition 
from the quantum to the classical regime occurs  precisely at {\bf the same point} $ r_A $ where
the chemical potential vanishes (see Fig. \ref{fnu}) showing the consistency and power of our treatment. This point defines the transition
from the quantum to the classical behaviour.}
\label{R}
\end{figure}

Inserting the phase-space density eq.(\ref{RTF}) into eq.(\ref{defR}) yields for the ratio  $ {\cal R} (r) $,
\be\label{defRa}
{\cal R} (r)  = 2 \; \sqrt5 \; \left( \frac{\pi}{81} \right)^{\! \! \frac13} \; 
\frac{I_2^{\frac{5}{6}}(\nu(\xi))}{I_4^{\frac12}(\nu(\xi))} = 1.513805 \; \frac{I_2^{\frac{5}{6}}(\nu(\xi))}{I_4^{\frac12}(\nu(\xi))} \; .
\ee
In Fig. \ref{R} we plot $ \log_{10} \cal R $ vs. $ \log_{10} (r/r_h) $ for the three representative galaxy solutions.
{\vskip 0.2cm}

Comparing now Figs.\ref{fnu} and \ref{R} we see  that $ \nu(\xi) $ {\bf changes sign} indicating the transition
from the quantum to the classical galaxy regime {\bf precisely  at the same point} where $ {\cal R} \simeq 1 $,
as it must be. This result shows the power and consistency of our treatment.

\section{Systematic study of the Thomas-Fermi Galaxy solutions with a central supermassive black hole}

We present in this section our extensive study of the Thomas-Fermi Galaxy solutions with a 
central supermassive black hole.

{\vskip 0.2cm}

As stated in subsection \ref{locteq}, each galaxy solution with a central black hole depends only on {\bf two
free parameters}:  $ \xi_0 $  and $ A $ defining the boundary conditions near the center [see eq.(\ref{conbh})], $ \xi_0$ being the dimensionless central radius and $A$ characterizing the central chemical potential behaviour. 

{\vskip 0.2cm}

\begin{figure}
\begin{turn}{-90}
\psfrag{"Camh.dat"}{No Black hole}
\psfrag{"Camhxicero001.dat"}{$ 0 < \xi_0 < 0.3 $}
\psfrag{"Camhxicero01.dat"}{$ \xi_0 = 0.1 $}
\psfrag{"Camhxicero03.dat"}{$ \xi_0 = 0.3 $}
\psfrag{"Camhxicero1.dat"}{$ \xi_0 = 1 $}
\psfrag{"Camhxicero2.dat"}{$ \xi_0 = 2 $}
\psfrag{"Camhxicero5.dat"}{$ \xi_0 = 5 $}
\psfrag{"Camhxicero7.dat"}{$ \xi_0 = 7 $}
\psfrag{"Camhxicero10.dat"}{$ \xi_0 = 10 $}
\psfrag{"Camhxicero12.dat"}{$ \xi_0 = 12 $}
\psfrag{"Camhxicero60.dat"}{$ \xi_0 = 60 $}
\psfrag{"Camhxicero70.dat"}{$ \xi_0 = 70 $}
\psfrag{"trucho.dato"}{}
\psfrag{log M halo}{$ \log_{10} M_h/M_\odot $}
\includegraphics[height=12.cm,width=10.cm]{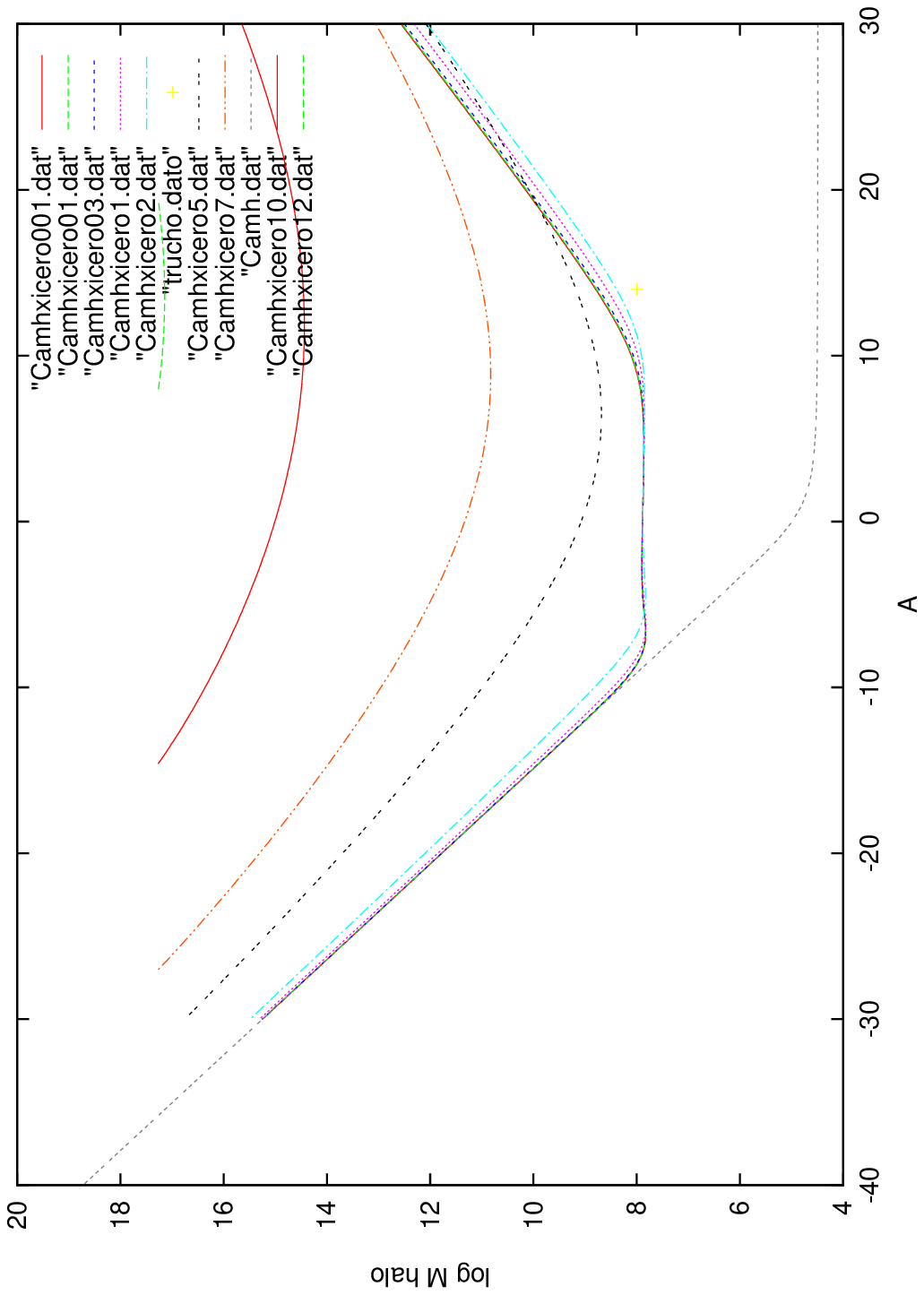}
\end{turn}
\caption{The halo mass $ \log_{10} M_h $ vs.  the constant $ A $ of the chemical potential behaviour at the origen for fixed values of $ \xi_0 $.           
The halo mass $ M_h $ increases with $ \xi_0 $ at fixed $ A $. $ M_h $ increases
when the absolute value of $ A $ increases at fixed $ \xi_0 > 0 $.
In absence of the central black hole, the halo mass monotonically decreases when $ A $ increases till
$ M_h $ reaches its {\it minimal value}  eq.(\ref{mhmin})
 at the degenerate quantum limit {\it at zero temperature} \cite{newas,astro,eqesta}.
In the presence of a central black hole, we find a {\it larger} minimal value for the halo mass $M_h^{min}$  eq.(\ref{bhmhmin}) with a {\it  non zero minimal temperature} $T_0^{min}$ eq.(\ref{bhtmin}) .
Therefore, there is an {\bf important qualitative} difference between galaxy solutions with a black hole
$ \xi_0 > 0 $, and galaxy solutions without a black hole $ \xi_0 = 0 $.}
\label{amh}
\end{figure}

We plot in Fig. \ref{amh} the halo mass $ \log_{10} M_h $ vs. $ A $ for fixed values of $ \xi_0 $.

We see that the halo mass $ M_h $ increases with $ \xi_0 $ at fixed $ A $. In addition, at fixed $ \xi_0 > 0 $,
$ M_h $ increases when the absolute value of $ A $ increases .

{\vskip 0.2cm}

There is an {\bf important qualitative} difference between galaxy solutions with a black hole
($ \xi_0 > 0 $ ), and galaxy solutions without a black hole ($ \xi_0 = 0 $).
In the absence of the central black hole, the halo mass $ M_h $ monotonically decreases when $ A $ increases till
$ M_h $ reaches a minimal value which is the degenerate quantum limit at zero temperature \cite{newas}, \cite{astro}, \cite{eqesta}:
\be\label{mhmin}
 M_h^{min} = 3.0999 \; 10^4 \; \left(\frac{2 \, {\rm keV}}{m}\right)^{\! \! \frac{16}5} \;
 \left(\frac{\Sigma_0 \;  {\rm pc}^2}{120 \; M_\odot}\right)^{\! \! \frac35}  \;  M_\odot 
, \; \;T_0^{min} = 0, \; {\rm {\bf without ~ central ~ black ~ hole}} \; .
\ee

In the presence of a central black hole, we find that the halo mass takes as minimal value

\be\label{bhmhmin}
 M_h^{min} =  6.892 \; 10^7 \; \left(\frac{2 \, {\rm keV}}{m}\right)^{\! \! \frac{16}5} \;
 \left(\frac{\Sigma_0 \;  {\rm pc}^2}{120 \; M_\odot}\right)^{\! \! \frac35}  \;  M_\odot 
\;, \quad {\rm {\bf with ~ central ~ black ~ hole}} \; .
\ee

This situation is clearly shown in Fig. \ref{amh}. The value of $M_h^{min}$ with a central black hole is $2.2233 \; 10^3$ times larger than without the black hole.
Notice that the small galaxy solution eq.(\ref{res3gal}) is just 11 \% larger in halo mass than the 
minimal galaxy eq.(\ref{bhmhmin}) with central black hole.

{\vskip 0.1cm}

We conclude that galaxies possesing a central black hole are in the dilute Boltzmann regime because of their large mass $ M_h >  M_h^{min} $ \cite{eqesta}. In addition, compact galaxies with $ M_h < M_h^{min} $, in particular
ultracompact galaxies in the quantum regime $ M_h < 2.3 \; 10^6  \;  M_\odot $ \cite{eqesta},
{\bf cannot} harbor central black holes. 

{\vskip 0.2cm}

\begin{figure}
\begin{turn}{-90}
\psfrag{"Cat0.dat"}{No Black hole}
\psfrag{"Cat0xicero001.dat"}{$ 0 < \xi_0 < 0.3 $}
\psfrag{"Cat0xicero01.dat"}{$ \xi_0 = 0.1 $}
\psfrag{"Cat0xicero03.dat"}{$ \xi_0 = 0.3 $}
\psfrag{"Cat0xicero1.dat"}{$ \xi_0 = 1 $}
\psfrag{"Cat0xicero2.dat"}{$ \xi_0 = 2 $}
\psfrag{"Cat0xicero5.dat"}{$ \xi_0 = 5 $}
\psfrag{"Cat0xicero7.dat"}{$ \xi_0 = 7 $}
\psfrag{"Cat0xicero10.dat"}{$ \xi_0 = 10 $}
\psfrag{"Cat0xicero12.dat"}{$ \xi_0 = 12 $}
\psfrag{"Cat0xicero60.dat"}{$ \xi_0 = 60 $}
\psfrag{"Cat0xicero70.dat"}{$ \xi_0 = 70 $}
\psfrag{"truch.dat"}{}
\psfrag{log M halo}{$ \log_{10} M_h/M_\odot $}
\includegraphics[height=12.cm,width=10.cm]{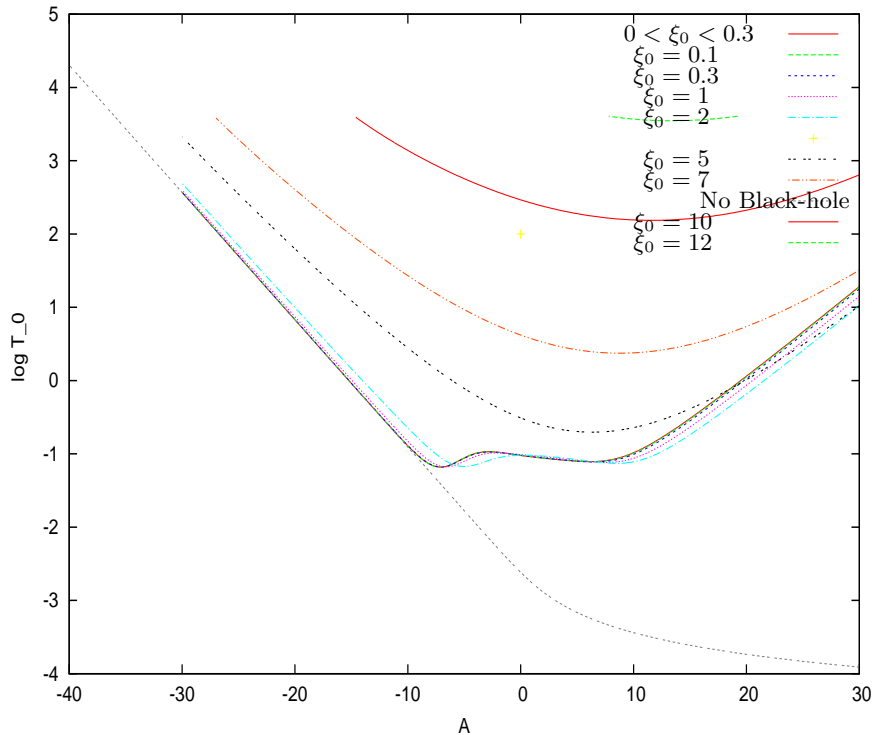}
\end{turn}
\caption{The galaxy temperature $ \log_{10} (T_0/{\rm K}) $ vs. the constant $ A $ of the chemical potential behaviour at the origin, for fixed values of $ \xi_0 $.
As for the halo mass $ M_h $, the galaxy temperature $ T_0 $ increases with $ \xi_0 $ at fixed $ A $. $ T_0 $
increases when the absolute value of $ A $ increases at fixed $ \xi_0 > 0 $.
In the absence of a black hole, the galaxy temperature $ T_0 $ tends to zero for $ A \to \infty $ (at the exact Fermi degenerate state) while in the presence o f a central black hole,  we find that $ T_0 $ is always {\it larger} than a {\it minimal non-zero value} $ T_0^{min} $ given by eq.(\ref{bhtmin}).}
\label{at0}
\end{figure}

We plot in Fig. \ref{at0} the galaxy temperature $ \log_{10} T_0/{\rm K} $ vs.   the characteristic central chemical potential constant $ A $ for fixed values of $ \xi_0 $.

Similarly to the halo mass $ M_h $, the galaxy temperature $ T_0 $ increases with $ \xi_0 $ at fixed $ A $. On the other hand, at fixed $ \xi_0 > 0 $, $ T_0 $ increases when the absolute value of $ A $ increases.

In the absence of a black hole, the galaxy temperature $ T_0 $ tends to zero for  $ A \to \infty $,
while in the presence of a central black hole   we find that $ T_0 $ is always {\it larger} than the {\it non zero minimal value}: 

\be \label{bhtmin}
T_0^{min} = 0.06928 \;  \;  \left(\frac{2 \, {\rm keV}}{m}\right)^{\! \! \frac35} \; 
\left(\frac{\Sigma_0 \;  {\rm pc}^2}{120 \; M_\odot}\right)^{\! \! \frac45}  \; {\rm K} \; 
, \quad {\rm {\bf with ~ central ~ black ~ hole}} \; .
\ee

\medskip

The presence of the supermassive black hole {\bf
heats-up} the dark matter gas
and prevents it to become an exact degenerate gas at
zero temperature. The minimal mass and size and most
compact galaxy state with a supermassive black hole is
a nearly degenerate state at very low temperature as
seen from eq. (\ref{bhtmin}).

\medskip

The mass of the supermassive black hole $ M_{BH} $
monotonically increases with $ \xi_0 $ at fixed $ A $.
In addition, for  $ \xi_0 < 0.3 $, that is for small
supermassive black holes,  and all $ A $, the galaxy
parameters as halo mass $ M_h $, halo radius $ r_h $,
virial mass $ M_{vir} $ and galaxy temperature $ T_0 $
become {\bf independent} of $ \xi_0 $ showing {\it a
limiting galaxy solution}. Only the BH mass
depends on  $ \xi_0 $ in this regime.

\medskip

\begin{figure}
\begin{turn}{-90}
\psfrag{"Cmbhmhxicero001.dat"}{$ \xi_0 = 0.01 $}
\psfrag{"Cmbhmhxicero01.dat"}{$ \xi_0 = 0.1 $}
\psfrag{"Cmbhmhxicero03.dat"}{$ \xi_0 = 0.3 $}
\psfrag{"Cmbhmhxicero1.dat"}{$ \xi_0 = 1 $}
\psfrag{"Cmbhmhxicero2.dat"}{$ \xi_0 = 2 $}
\psfrag{"Cmbhmhxicero5.dat"}{$ \xi_0 = 5 $}
\psfrag{"Cmbhmhxicero7.dat"}{$ \xi_0 = 7 $}
\psfrag{"Cmbhmhxicero10.dat"}{$ \xi_0 = 10 $}
\psfrag{"Cmbhmhxicero12.dat"}{$ \xi_0 = 12 $}
\psfrag{"Cmbhmhxicero60.dat"}{$ \xi_0 = 60 $}
\psfrag{"Cmbhmhxicero65.dat"}{$ \xi_0 = 65 $}
\psfrag{"trucho.dato"}{}
\psfrag{log M halo}{$ \log_{10} M_h/M_\odot $}
\psfrag{log M BH}{$ \log_{10} M_{BH}/M_\odot $}
\includegraphics[height=12.cm,width=10.cm]{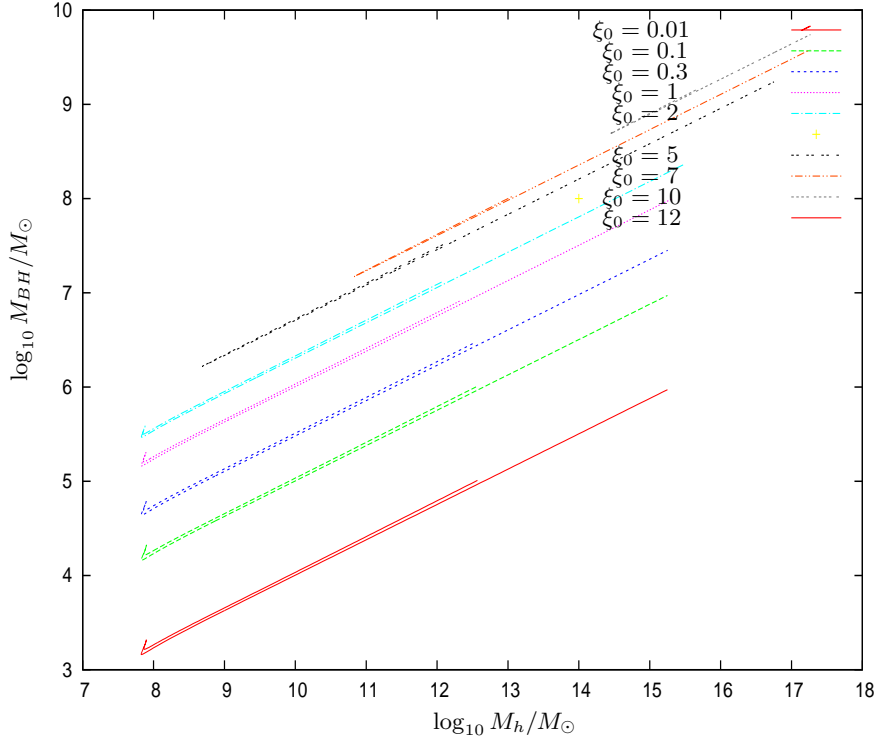}
\end{turn}
\caption{The black hole mass $ \log_{10} M_{BH} $ vs. the halo mass $ \log_{10} M_h $. The black hole mass 
$ M_{BH} $ turns out to be a {\bf two-valued} function of $ M_h $. For each value of $ M_h $ there are 
two values for $ M_{BH} $. These two values of $ M_{BH} $ for a given $ M_h $ 
are quite close to each other. This two-valued dependence on $ M_h $ is a direct consequence
of the dependence of $ M_h $ on the  central chemical potential behaviour characterized by the constant $ A $  as shown in Fig. \ref{amh}.}
\label{bhmh}
\end{figure}

We depict in Fig. \ref{bhmh} the black hole mass $
\log_{10} M_{BH} $ vs. the halo mass $ \log_{10} M_h
$.
We see that $ M_{BH} $ is a {\bf two-valued} function
of $ M_h $. For each value of $ M_h $ there are 
two possible values for $ M_{BH} $. These two values
of $ M_{BH} $ for a given $ M_h $ 
are quite close to each other. This two-valued
dependence on $ M_h $ is a direct consequence
of the dependence of $ M_h $ on $ A $ shown in Fig.
\ref{amh}.

{\vskip 0.1cm}

We see at the branch-points on the left in Fig.
\ref{bhmh},  the minimal galactic halo mass $
M_h^{min} $ 
eq.(\ref{bhmhmin}) when a supermassive black hole is
present.

{\vskip 0.2cm}

At {\bf fixed} $ \xi_0 $, as shown in Fig. \ref{bhmh},
the central black hole mass $ M_{BH} $ scales with the
halo mass $ M_h$ as
$$
 M_{BH} = D(\xi_0) \;  M_h^\frac38 \; ,
$$
where $ D(\xi_0) $ is an increasing function of $ \xi_0 $.

\begin{figure}
\begin{turn}{-90}
\psfrag{"Cmht0.dat"}{No Black hole}
\psfrag{"Cmht0xicero001.dat"}{$ 0 < \xi_0 < 0.3 $}
\psfrag{"Cmht0xicero01.dat"}{}
\psfrag{"Cmht0xicero03.dat"}{$ \xi_0 = 0.3 $}
\psfrag{"Cmht0xicero5.dat"}{$ \xi_0 = 5 $}
\psfrag{"Cmht0xicero10.dat"}{$ \xi_0 = 10 $}
\psfrag{"Cmht0xicero1.dat"}{$ \xi_0 = 1 $}
\psfrag{"Cmht0xicero2.dat"}{$ \xi_0 = 2 $}
\psfrag{"Cmht0xicero7.dat"}{$ \xi_0 = 7 $}
\psfrag{"Cmht0xicero12.dat"}{$ \xi_0 = 12 $}
\psfrag{"Cmht0xicero50.dat"}{$ \xi_0 = 50 $}
\psfrag{"Cmht0xicero60.dat"}{$ \xi_0 = 60 $}
\psfrag{"Cmht0xicero65.dat"}{$ \xi_0 = 65 $}
\psfrag{"Cmht0xicero70.dat"}{$ \xi_0 = 70 $}
\psfrag{"trucho.dat"}{}
\psfrag{log M halo}{$ \log_{10} M_h/M_\odot $}
\psfrag{log T_0}{$ \log_{10} T_0/{\rm K} $}
\includegraphics[height=12.cm,width=10.cm]{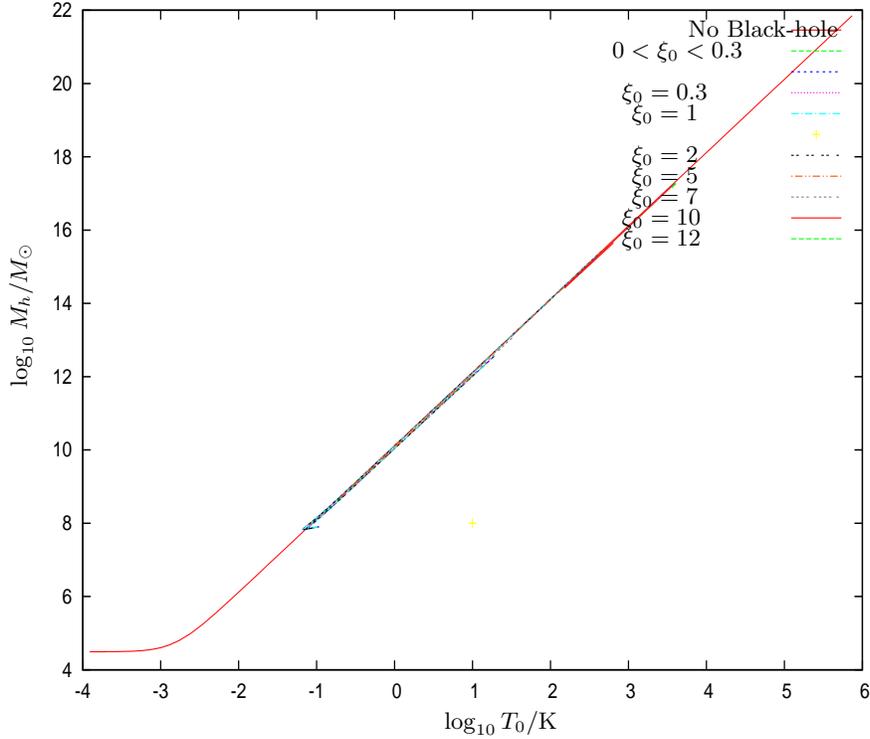}
\end{turn}
\caption{The halo galaxy mass $ \log_{10} M_h $ vs. the galaxy temperature $ \log_{10} T_0/{\rm K} $.
$ M_h $ turns to be a {\bf two-valued} function of $ T_0 $.
The halo mass $ M_h $ grows when $ T_0 $ increases. Colder galaxies are smaller while warmer galaxies are larger.
We see at the branch-points in Fig. \ref{mht0} the minimal galaxy temperature $ T_0^{min}  $
eq.(\ref{bhtmin}) when a supermassive black hole is present.}
\label{mht0}
\end{figure}

\begin{itemize}
\item {We plot in Fig. \ref{mht0} the halo galaxy mass $
\log_{10} M_h $ vs. 
the galaxy temperature $ \log_{10} T_0/{\rm K} $.
The halo mass $ M_h $ grows when  $ T_0 $ increases.
Colder galaxies are smaller. Warmer galaxies are
larger.

We see  at the branch-points in Fig. \ref{mht0} the
minimal galaxy temperature $ T_0^{min} $
eq.(\ref{bhtmin}) when a supermassive black hole is
present.}

\item{We find galaxy solutions with central black holes for
arbitrarily small values $ \xi_0 > 0 $ and
correspondingly arbitrarily small central BH mass.
There is no emergence of a minimal  mass for the central black hole.}
\end{itemize}

\begin{figure}
\begin{turn}{-90}
\psfrag{"Cmhrh.dat"}{No Black hole}
\psfrag{"Cmhrhxicero001.dat"}{$ 0 < \xi_0 < 0.3 $}
\psfrag{"Cmhrhxicero1.dat"}{$ \xi_0 = 1 $}
\psfrag{"Cmhrhxicero01.dat"}{}
\psfrag{"Cmhrhxicero03.dat"}{$ \xi_0 = 0.3 $}
\psfrag{"Cmhrhxicero2.dat"}{$ \xi_0 = 2 $}
\psfrag{"Cmhrhxicero7.dat"}{$ \xi_0 = 7 $}
\psfrag{"Cmhrhxicero10.dat"}{$ \xi_0 = 10 $}
\psfrag{"Cmhrhxicero5.dat"}{$ \xi_0 = 5 $}
\psfrag{"Cmhrhxicero12.dat"}{$ \xi_0 = 12 $}
\psfrag{"Cmhrhxicero40.dat"}{$ \xi_0 = 40 $}
\psfrag{"trucho.dato"}{}
\psfrag{log M halo}{$ \log_{10} M_h/M_\odot $}
\psfrag{log rh}{$ \log r_h/{\rm pc} $}
\includegraphics[height=12.cm,width=10.cm]{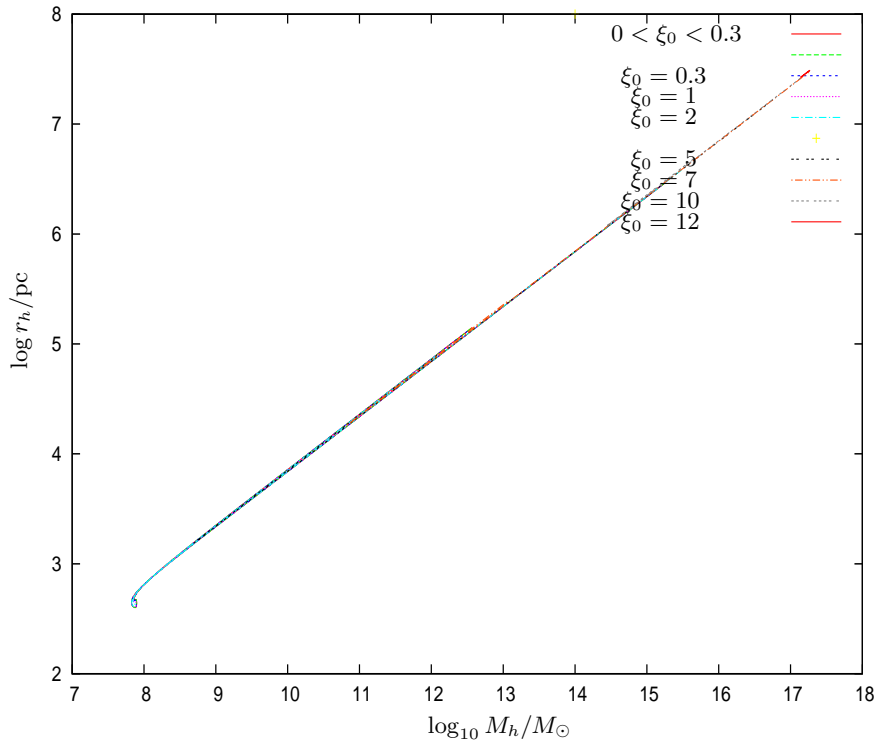}
\end{turn}
\caption{The halo radius $ \log_{10} r_h $ vs. 
the common logarithm of the halo mass $ \log_{10} M_h $ for galaxies with
supermassive central black holes of many different masses. 
$  r_h $ turns to be a {\bf two-valued} function of $ M_h $. We see that $ M_h $ accurately
scales as $ r_h^2 $.
The same scaling was found in the Thomas-Fermi approach for galaxies
in the absence of black holes \cite{newas,astro,eqesta}.}
\label{mhrh}
\end{figure}

\begin{figure}
\begin{turn}{-90}
\psfrag{"Cmhrhcoef.dat"}{No Black hole}
\psfrag{"Cmhrhcoefxicero001.dat"}{$ 0 < \xi_0 < 0.3 $}
\psfrag{"Cmhrhcoefxicero01.dat"}{}
\psfrag{"Cmhrhcoefxicero03.dat"}{}
\psfrag{"Cmhrhcoefxicero7.dat"}{$ \xi_0 = 7 $}
\psfrag{"Cmhrhcoefxicero10.dat"}{$ \xi_0 = 10 $}
\psfrag{"Cmhrhcoefxicero1.dat"}{$ \xi_0 = 1 $}
\psfrag{"Cmhrhcoefxicero2.dat"}{$ \xi_0 = 2 $}
\psfrag{"Cmhrhcoefxicero5.dat"}{$ \xi_0 = 5 $}
\psfrag{"Cmhrhcoefxicero12.dat"}{$ \xi_0 = 12 $}
\psfrag{"Cmhrhcoefxicero60.dat"}{$ \xi_0 = 60 $}
\psfrag{"Cmhrhcoefxicero70.dat"}{$ \xi_0 = 70 $}
\psfrag{"truchi.dat"}{}
\psfrag{log M halo}{$ \log_{10} M_h/M_\odot $}
\psfrag{b = Mh/[Sigma0 rh^2]}{$ b \equiv M_h/[\Sigma_0 \; r_h^2] $}
\includegraphics[height=12.cm,width=10.cm]{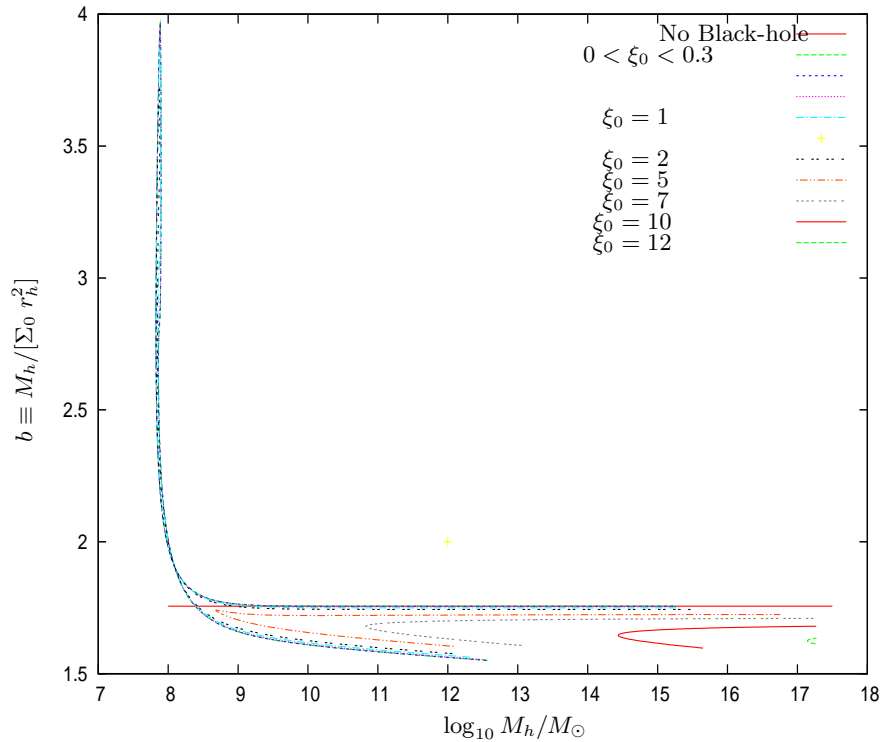}
\end{turn}
\caption{ The scaling amplitude $ b \equiv M_h/[\Sigma_0 \; r_h^2]$ as a function of the halo mass $ M_h $. 
Except for halo masses near the minimum
halo mass $ M_h^{min} $ eq.(\ref{bhmhmin}),  $ b $ in the presence of a central black hole takes values up to
10\% below its value  $ 1.75572 $ in the absence of a central black hole eq.(\ref{bsinbh}).
The continuous red horizontal line $ b = 1.75572 $ corresponds to galaxies without central
black holes [eq.(\ref{bsinbh})].}
\label{coef}
\end{figure}

\begin{figure}
\begin{turn}{-90}
\psfrag{"Cmbhrh.dat"}{No Black hole}
\psfrag{"Cmbhrhxicero001.dat"}{$ 0 < \xi_0 < 0.3 $}
\psfrag{"Cmbhrhxicero01.dat"}{$ \xi_0 = 0.1 $}
\psfrag{"Cmbhrhxicero7.dat"}{$ \xi_0 = 7 $}
\psfrag{"Cmbhrhxicero10.dat"}{$ \xi_0 = 10 $}
\psfrag{"Cmbhrhxicero03.dat"}{$ \xi_0 = 0.3 $}
\psfrag{"Cmbhrhxicero1.dat"}{$ \xi_0 = 1 $}
\psfrag{"Cmbhrhxicero2.dat"}{$ \xi_0 = 2 $}
\psfrag{"Cmbhrhxicero5.dat"}{$ \xi_0 = 5 $}
\psfrag{"Cmbhrhxicero12.dat"}{$ \xi_0 = 12 $}
\psfrag{"Cmbhrhxicero70.dat"}{$ \xi_0 = 70 $}
\psfrag{"trufa.dat"}{}
\psfrag{log M BH}{$ \log_{10} M_{BH}/M_\odot $}
\psfrag{log rh}{$ \log r_h/{\rm pc} $}
\includegraphics[height=12.cm,width=10.cm]{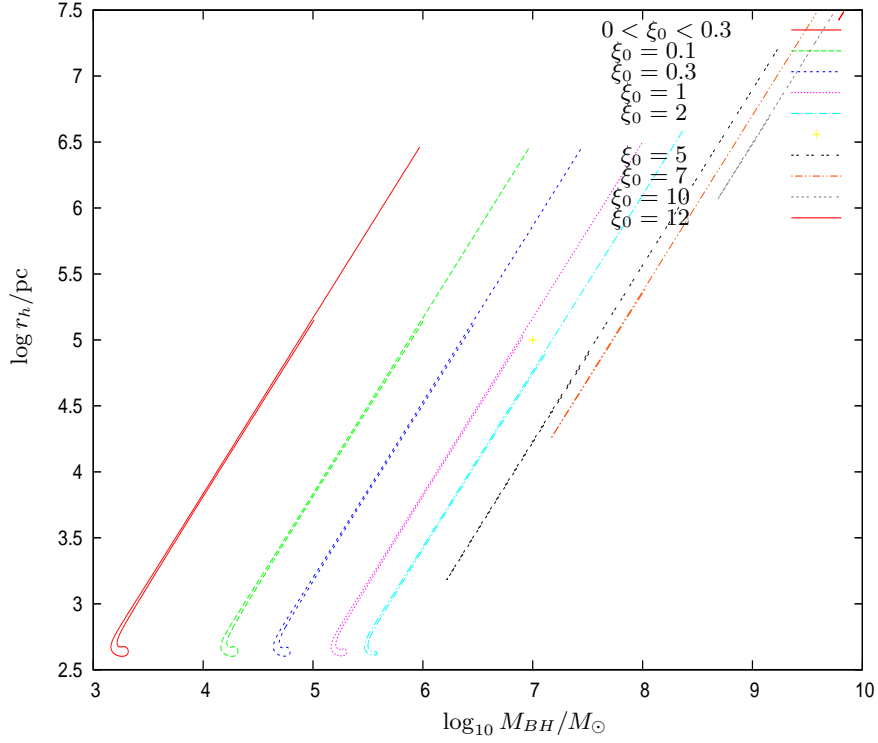}
\end{turn}
\caption{The common logarithm of the halo radius $ \log_{10} r_h $ vs. 
the common logarithm of the central black hole mass $ \log_{10} M_{BH} $ for many
galaxy solutions. The halo radius $ r_h $ turns to be a {\bf double-valued function} of $ M_{BH} $.
Remarkably, $ r_h $ scales with the black hole mass for {\bf fixed} $ \xi_0 $ as  $ r_h = C(\xi_0) \;  M_{BH}^\frac43 $ where the constant $ C(\xi_0) $ is a decreasing function of $ \xi_0 $.}
\label{mbhrh}
\end{figure}

\begin{figure}
\begin{turn}{-90}
\psfrag{Log10 P}{$ \log_{10} P(r) $}
\psfrag{"Zprchi.dat"}{Small Galaxy}
\psfrag{"Zprmed.dat"}{Medium Galaxy}
\psfrag{"Zprgran.dat"}{Large Galaxy}
\psfrag{x = log xi/xh}{$ \log_{10} r/r_h $}
\psfrag{ACA xa}{$ x_A \downarrow $}
\psfrag{ACA xi}{$ x_i \downarrow $}
\psfrag{ACA xh}{$ x_h \downarrow $}
\includegraphics[height=12.cm,width=10.cm]{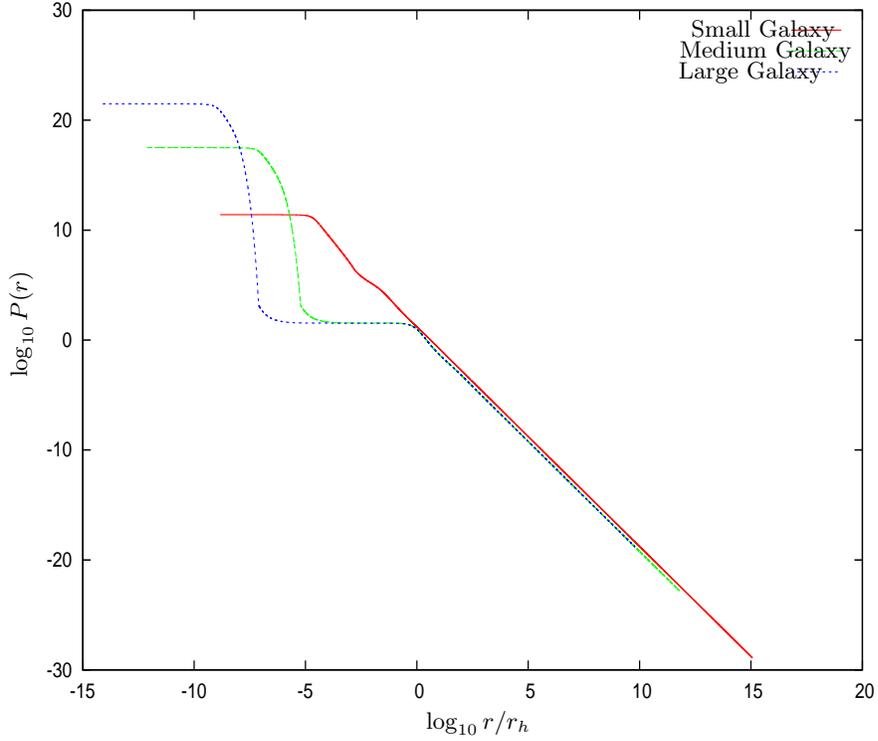}
\end{turn}
\caption{The logarithm of the local pressure $ \log_{10} P(r) $ vs. $ \log_{10} (r/r_h) $ for the three
galaxy solutions with central SMBH. Notice the huge values of $ P(r) $ in the quantum (high density) region $ r < r_A $
and its sharp decrease entering the classical (dilute) region $ r > r_A $.}
\label{pr}
\end{figure}

\begin{figure}
\begin{turn}{-90}
\psfrag{Log10 P}{$ \log_{10} P(r) $}
\psfrag{"Zprhochi.dat"}{Small Galaxy}
\psfrag{"Zprhomed.dat"}{Medium Galaxy}
\psfrag{"Zprhogran.dat"}{Large Galaxy}
\psfrag{Log10 rho/rho0}{$ \log_{10} \rho(r)/ \rho_0 $}
\psfrag{ACA xa}{$ x_A \downarrow $}
\psfrag{ACA xi}{$ x_i \downarrow $}
\psfrag{ACA xh}{$ x_h \downarrow $}
\includegraphics[height=12.cm,width=10.cm]{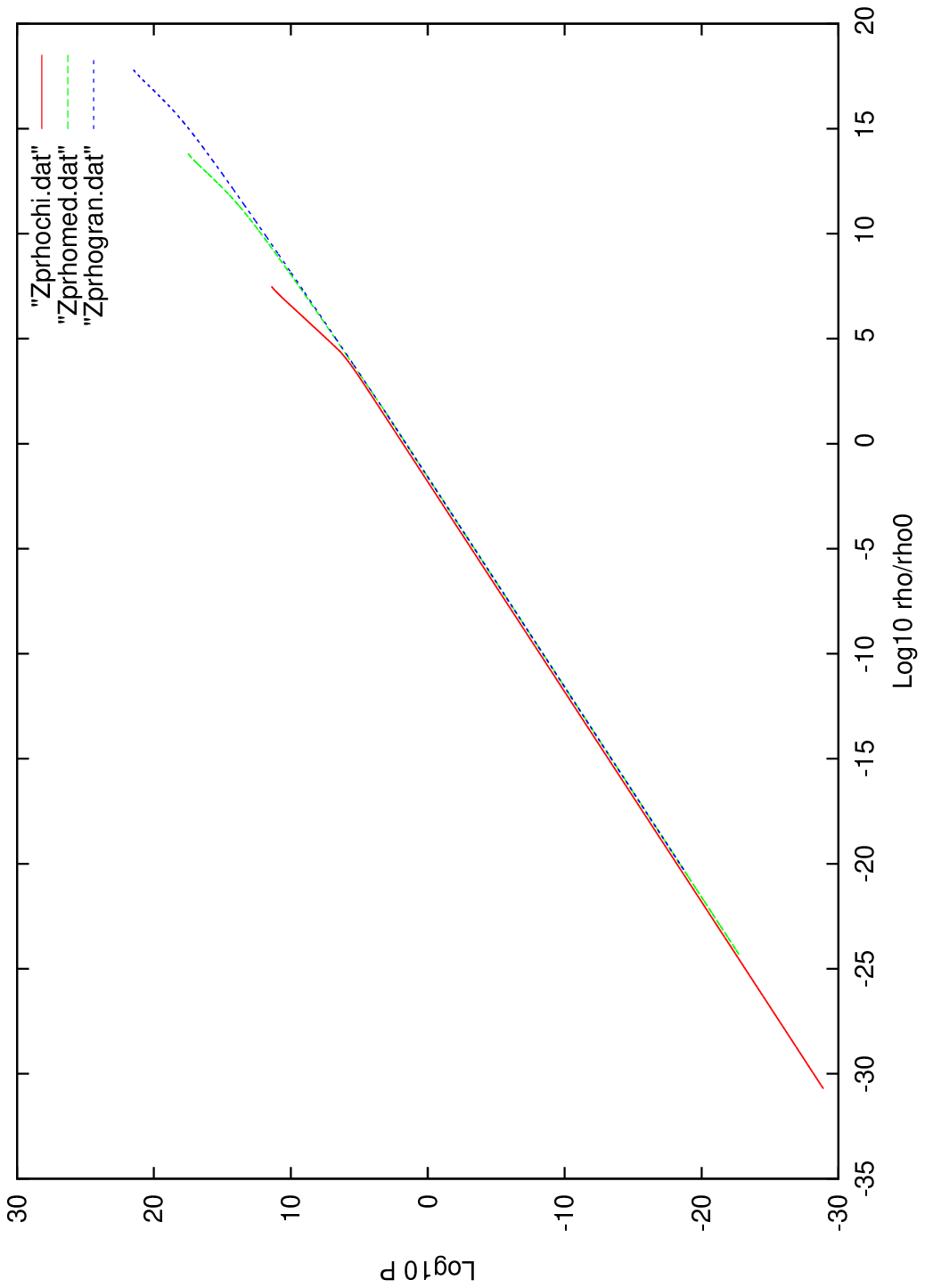}
\end{turn}
\caption{The obtained equation of state of the galaxy plus central SMBH system: The logarithm of the local pressure $ \log_{10} P(r) $ vs. 
$ \log_{10} \rho(r)/ \rho_0 $. In all the cases we find almost straight lines of unit slope. The equation of state is
a perfect gas equation of state in the Boltzmann classical region. In the quantum gas (dense) region the equation of state
becomes steeper than the perfect gas. Galaxies with central black holes 
are in the dilute Boltzmann regime because their halo masses are
$ M_h >  M_h^{min} $ eq.(\ref{bhmhmin}). This explains the perfect gas equation of state.}
\label{prho}
\end{figure}

\subsection{Universal Scaling relations in the presence of central black holes}

We plot in Fig. \ref{mhrh} the ordinary logarithm of the halo radius $ \log_{10} r_h $ vs. 
the ordinary logarithm of the halo mass $ \log_{10} M_h $ for galaxies with
central black holes of many
different masses. We see in {\bf all cases} that $ M_h $ scales as $ r_h^2 $.
The same scaling was found in the Thomas-Fermi approach to galaxies
in absence of black holes \cite{newas}, \cite {astro}, \cite{eqesta}.

\medskip

The halo mass in the absence of a central black hole behaves in
the Thomas-Fermi approach as \cite{eqesta}
\be \label{bsinbh}
 M_h = 1.75572 \; \Sigma_0 \; r_h^2 \quad ,\quad {\rm {\bf without ~ central ~ black ~ hole}} \; .
\ee
The proportionality factor in this scaling relation is confirmed by the galaxy data \cite{eqesta}.

\medskip

In the presence of a central black hole we find in the Thomas-Fermi approach 
an analogous relation
\be\label{bbh}
 M_h = b \; \; \Sigma_0 \; r_h^2 \quad ,\quad {\rm {\bf with ~ central ~ black ~ hole}} \; ,
\ee
where the coefficient $ b $ turns to be of order unity. 

\medskip

We plot in Fig. \ref{coef}
the coefficient $ b $ as a function of the halo mass $ M_h $. We see that except for halo masses near the minimum
halo mass $ M_h^{min} $ ,  $ b $ in the presence of a central black hole takes values up to
10\% below its value in the absence of a central black hole eq.(\ref{bsinbh}).
For halo masses near $ M_h^{min} , \; b $ increases reaching values $ b \leq 4 $.
For very large halos and central black holes, $b$ could be as small as about $1.6$.

{\vskip 0.2cm}

That is, the coefficient $ b $ changes at most by a factor from $ 1/2 $ up to $ 2 $
while the halo mass $ M_h $ varies ten orders of magnitude.
As shown by Fig.\ref{coef}, the coefficient $ b $ turns to be a two-valued function of $ M_h $.

{\vskip 0.2cm}

The coefficient $ b $ turns to be independent of the precise value of the WDM particle mass $ m $.
This is due to the fact that the scaling relation eq.(\ref{bbh}) as well as eq.(\ref{bsinbh})
apply in the classical Boltzmann regime of the galaxy.

{\vskip 0.2cm}

In summary, the scaling relation eq.(\ref{bbh}) and the coefficient $ b $ turn  out to be remarkably robust.

{\vskip 0.2cm}

We plot in Fig. \ref{mbhrh} the ordinary logarithm of the halo radius $ \log_{10} r_h $ versus 
the ordinary logarithm of the central black hole mass $ \log_{10} M_{BH} $ for many
galaxy solutions. The halo radius $ r_h $ turns to be a double-valued function of $ M_{BH} $.
{\bf Remarkably}, $ r_h $ scales for {\bf fixed} $ \xi_0 $ as  
\be
r_h = C(\xi_0) \;  M_{BH}^\frac43 .
\ee
The constant $ C(\xi_0) $ turns out to be a decreasing function of $ \xi_0 $.

\subsection{Pressure and equation of state in the presence of central black holes}

The local pressure $ P(r) $ is given by eq. (\ref{presion}). In Fig. \ref{pr} we plot $ \log_{10} P(r) $ vs. $ \log_{10} (r/r_h) $ for the three representative galaxy solutions. We see that $ P(r) $ monotonically
decreases with $ r $. The pressure $ P(r) $ takes huge values in the quantum (high density) region $ r < r_A $ and then
it sharply decreases entering the classical (dilute) region $ r > r_A $.

{\vskip 0.2cm}

In Fig. \ref{prho}  we plot  $ \log_{10} P(r) $ vs. $ \log_{10} \rho(r)/ \rho_0 $ for the three galaxy solutions with central SMBH. 
We see that the three curves almost coincide and that
they are almost straight lines of unit slope. That is, the equation of state is
in very good approximation a perfect gas equation of state.  This perfect gas equation of state
stems from the fact that galaxies with central black holes have halo masses
$ M_h >  M_h^{min} $ eq.(\ref{bhmhmin}) and therefore belong to the {\it dilute }
Boltzmann classical regime \cite{eqesta}. The equation of state turns out a local ($r$-dependent) perfect gas equation of state because of the gravitational interaction, (WDM self-gravitating perfect gas). 

{\vskip 0.2cm}

Indeed, for galaxies with central black holes the WDM is in a quantum (highly compact) regime
inside the quantum radius $ r_A $. However, because $ r_A $ is in the parsec scale or smaller [see eqs.(\ref{res3gal})-(\ref{grangal})]
the bulk of the WDM is in the Boltzmann classical regime which consistently reflects in the perfect gas equation of state behaviour.

\section {Conclusions}

\begin{itemize}
\item{We have presented here a novel study of galaxies with
central supermassive black holes which shows itself
fruitful and enlighting. This framework stress the key role
of gravity and warm dark matter in structurating galaxies
with their central supermassive black holes and provides
correctly the major physical quantities to be first
obtained for the galaxy-black hole system: the masses,
sizes, densities, velocity dispersions, and their internal physical states. This also yields a physical and precise
characterisation of whether they are compact, ultracompact,
low density or large dilute galaxies, encompassed with 
their classical physics and quantum gas physical
properties.} 

\item{We thus found different regions structurating internally
the halo of the galaxy from the vicinity of the
supermassive central black hole region to the external
regions or virial radius. For all galaxy arboring a central
black hole there is a transition from the quantum to the
classical regime going from the more compact inner regions
which are in a quantum gas state till the classical dilute
regions in a Boltzmann-like state. This is accompanied by a
decreasing of the local temperature from the central warmer
regions to the colder external ones. The SMBH heats the DM
near around and prevents it to became exactly degenerated
at zero temperature. Although the inner DM quantum core is
highly compact in a nearly degenerate quantum gas state, it
is not at zero temperature. 
Inside $ r \lesssim 3 \; r_h $ the halo is thermalized at a
uniform or slowly varying local temperature $ T_0 $ which
tends to the circular temperature  $ T_c (r) $ at $ r \sim
3 \; r_h $.}

\item{We have formulated the problem of galaxy structure with
central supermassive black holes in the WDM Thomas-Fermi
approach and found the main physical magnitudes and
properties of the galaxy plus black hole system.  We solved
the corresponding equations and boundary conditions, find
three representative families of realistic galaxy solutions
(small, medium and large size galaxies) with central
supermassive black holes and provided a systematic analysis
of the new quantum and classical physics properties of the
system. The approach naturally incorporates the quantum
pressure of the self-gravitating dark matter fermions
showing its full power and clearness to treat the galaxy
plus supermassive black hole system. The realistic
astrophysical masses of supermassive black holes are
naturally obtained in this framework.}

\item{ We found the main important physical differences between
galaxies with and without the presence of a central black
hole. In the presence of a central black hole, both the
quantum and classical behaviours of the dark matter gas do
co-exist  generically in any galaxy from the compact small
galaxies to the dilute large ones and a novel galaxy halo
structure with three regions show up.} 

\item{The transition from the quantum to classical regime occurs
at the point $r_A$ where the chemical potential vanishes
and which is in addition, precisely and consistently, the
point where the particle wavelength and the interparticle
distance are equal (their ratio being a measure of the
quantum or classical properties of the system). The 
quantum radius $r_A $ is larger for the smaller and more
compact galaxies and diminishes with increasing galaxy and
black hole masses for the large dilute galaxies. The WDM
mass $ M_A $ inside the quantum galaxy radius  $ r_A $
represents only a small fraction of the halo mass $ M_h$ or
virial mass of the galaxy
but it is a significant fraction of the black hole mass $
M_{BH} $. $ M_A $ amounts
for 20\% of  $ M_{BH} $ for the medium and large galaxies
and 45\% for the small galaxies.}

\item{The minimal mass $M^{min}_h $ a galaxy should have in order
to harbor SMBHs have been found, which shows among other
features why compact or ultracompact galaxies (in the range
$ 10^4  \; M_\odot < M_h < 10^7 \; M_\odot$) cannot harbor
necessarily central black holes.} 

\item{Novel universal scaling relations in the presence of a
central supermassive black hole have been derived: black
hole mass $M_{BH}$ - halo radius $ r_h$ - halo mass $M_h$
relations. The black hole mass $ M_{BH} $  turns out to be
a two-valued function of the halo mass $ M_h$ and size
$r_h$, and we found the local pressure and equation of
state of the galaxy-black hole system and its different
regimes.} 

\item{A more detailed quantitative account of the main features and results of this paper is presented in the
Introduction-Section I.}

\item{ The circular velocities, galactic rotation curves in the WDM halo with central SMBH  are discussed, self-consistently computed and plotted in Section II, Eqs (2.46) to (2.49), Eqs (2.53)-(2.54) and Fig.4 of this paper together  with the obtained velocity dispersions. These results are presented for the three family of galaxy solutions with SMBHs obtained here with this approach: Small or Dwarf Galaxies, Medium Galaxies and Large Galaxies. They remarquably accompass the other relevant physical magnitudes obtained for these systems in this paper  with the same  approach.                              Towards the central regions, the circular velocity grows as in eq.(2.49) due to the central black hole field. As seen from Fig 4, the dispersion velocity  is constant in the Boltzmann (outer or classical) region and in the quantum (inner or compact) region, indicating WDM thermalization. For $r > r_h$, the circular velocity tends to the velocity dispersion.  Remarquably, this result confirms the same behaviour  we obtained independently with a different approach (the inverse problem or the Eddington integral equation for galaxies which we developped in Ref  \cite{eddi}), namely : given the observed  density profiles as input, the velocities, pressure and other galaxy magnitudes are obtained and analyzed as output. The observed density profiles being by definition real realistic data, the obtained results from them are trustable realistic magnitudes. Moreover, another robust verification of the keV WDM Thomas-Fermi approach are the 10 independent sets of observational data we used in Ref \citep{urc} for galaxy masses from $5 \; 10^9 M_\odot$ to $5 \; 10^{11} M_\odot$. The theoretical and observational rotation curves do agree. In addition, they agree extremely well with the observational rotation curves described by the empirical Burkert profile for $r \geq 2 r_h$. (they differ from each other by only $2.4$ per cent). These results show the success of the keV WDM  Thomas-Fermi approach to correctly describe the galaxy structures.}

\item{We have first investigated pure WDM galaxies with their
central black holes, because DM is on average the
over-dominant component in galaxies, and it is reasonable
then to investigate first the effects of gravity plus WDM.
This is thus a first approximation, more precisely the zero
order of a first approximation in which the visible matter
component, baryons, can be incorporated to provide a most
accurate and complete picture.  We have seen, that these
zero order results found here are already realist very good
and robust results and they set the basis and the direction
for improvements and a more complete understanding.}

\item{Baryons will provide corrections to this picture and will
allow to study other processes in which ordinary matter
naturally plays a role as the gas and star components, but
baryons will not change drastically the pure WDM results
found here which are the structural galaxy and black hole
properties, masses, sizes,  their scaling and  relations,
density profiles, the classical and quantum nature of the
halo regions and their physical, high density, medium
density or dilute state, the halo thermalization and virialization.} 

\item{This  predictive theory and the obtained classes of solutions include very well  the different  galaxy types through their generic and important physical quantitive properties as  the pressure, density,  equation of state, mass, halo structure, central black holes. Thus, we have primarily three  galaxy classes: large dilute galaxies, intermediate galaxies,  and small compact galaxies, whatever their astronomical empirical/historical name.  The {\it Milky Way} galaxy is one of the galaxies in the large dilute galaxy class we found with all the specific properties of this class, mass, structure and central SMBH. {\it Messier 87}  is a larger ("supergiant") galaxy within the large class of galaxies we found, and hosting  consequently, a bigger  central SMBH (M87).}

\item{As explained in the paper, the {\it central quantum WDM gaz}  is relevant  for the presence of the obtained  central non cusped cores and their correct sizes, and for the presence of the central SMBHs and their realistic mass values without any ad-hoc prescription.
Recall for instance Fig. 3 of the paper, which displays the density $\rho (r)$ normalized at the influence radius $r_i$, vs $r/r_h$ for the three family of galaxy solutions with central SMBHs we found: Large dilute galaxies, intermediate galaxies,  and small compact galaxies, covering the different types of  galaxies with their central SMBHs.  The {\it Milky Way} is within the Large dilute galaxy class we found with all the characteristic properties of this class:  mass, structure and central SMBH, namely $M_{BH} = 4.100 \; 10^6 M_{\odot}$,  galaxy mass $M = (0.8-1.5) \; 10^{12} M_{\odot}$ and $r_h = 580 +/- 120 kpc)$.
 Notice that in the {\it quantum WDM gas region} $r < r_A$,  the density is constant clearly exhibiting a {\it plateau} behaviour corresponding to the {\it quantum macroscopic}  Fermi DM gas behaviour in such region.
Fig. 3 shows that the local density behaviour is dominated by the black hole for $r \lesssim r_i$. Coherently, for $r_i \lesssim r \lesssim r_h$ the WDM gravitational field dominates over the black hole field and the galaxy core shows up. For medium and large galaxies as the Milky Way the core is seen as a {\it plateau}. At the same time the chemical potential is negative for $r > r_i > r_A$ and the WDM is a Boltzmann gas in this region.

The first or primary "signatures" are the set of galactic physical magnitudes and structural properties : sizes, masses, cored density profiles and their correct sizes. In particular,  {\it Dwarf} galaxies appear to be a {\it full quantum macroscopic system}. Dwarf galaxies are really interesting to observe in this respect, as tracers of the quantum keV WDM nature in nearly  degenerated states, their temperatures and properties.
These are important features all found and provided by the same and one single approach, without tailorated prescriptions, and without considering different approachs for each of the  different computed magnitudes.
Therefore, these are all "signatures" say for this approach.  

These results consistently accompass the ones shown in Fig.2 : the derivative of the chemical potential vs. $(r/rh)$ for the three families of galaxy solutions with central SMBH. For $r \lesssim r_i$ the behaviour is dictated by the central black hole.  For $r > r_i$, they are dominated by the WDM and in this region exhibit a similar behaviour to the Thomas-Fermi galaxy solutions without a central SMBH  \cite {newas},\cite {astro}, \cite {urc}, \cite{eqesta}. For galaxies with central black holes the WDM is in a quantum (highly compact) regime inside the quantum radius $r_A$. Because $r_A$ is in the parsec scale or smaller [see eqs.(3.8)-(3.10)] , the bulk of the WDM is then in the Boltzmann regime eg Fig. 13 and Fig. 14. In the quantum gas (dense) region the equation of state becomes steeper than the perfect gas. Notice the huge values of $P(r)$ in the quantum (high density) region $r < r_A$ and its sharp decrease entering the classical (dilute) region $r > r_A$ all consistent with the other results we found.}

 \item{ In all the obtained results, and in the Introduction we have 
 carefully compared  the results and solutions we obtained in this 
 paper for galaxy systems with a central black hole and without a 
 central black hole. From our results here we recover, in particular, 
 the galaxy structures, the cores of quantum WDM and their right sizes, 
 the velocity dispersions, the scaling relations, the equation of 
 state and the other related results  in the absence of black holes we already discussed in our previous works  eg Refs \cite {newas},\cite {astro},\cite {urc},\cite {eqesta},\cite {eddi} in which the 
 careful check for rotation curves, masses, scaling relations, 
 velocities, are in full agreement with observations 
 for the whole set of properties.  
 
Cored density profiles and their right size, halo masses, are in full 
agreement with the observations. The quantum DM nature in the central 
regions  is {\it not} an exotic property : It is the quantum nature of the 
degenerate or nearly degenerate gaz of DM particles. Interaction is 
fully gravitationnal, namely a self-gravitating and self-consistent  
WDM gaz. 
The first or primary " signatures " are therefore the set of galactic 
physical magnitudes and structural properties: sizes, masses , 
central cored profiles, velocity distributions, surface density we 
have found and confronted to real astronomical observations.  Other 
effects as the influence such DM structures could in turn exerce on 
the propagation of generated gravitational waves, or on the accretion 
processes, are superimpossed effects,  or secondary dark matter 
processes or secondary signatures, a problem which would require to  
be analysed by its own,  and which study is clearly beyond the scope 
of the present  paper which is devoted to the primary dark matter 
effects, namely the dark matter galactic structures. Those secondary 
effects as the orbits, diffusion and 
absorption in the different regions and regimes around the BH and 
requiring the interaction in propagation with other non dark matter 
components, as electromagnetic effects and accretion plasmas are not 
the subject of this paper.}

\item{ For the primary objectives of obtaining  the galaxy structural magnitudes : eg the realistic astrophysical masses of the galaxies, the realistic SMBH central masses, their sizes, velocities, cored density  and pressure profiles, the newtonian treatment is largely enough.  Recall that the Thomas-Fermi approach is a statistical many body approach. Near the black hole horizon, there will show up effects of  spiraling, orbiting,  or a glory effect (180 degrees back scattering) but it does not affect really the properties and magnitudes of the galaxy-black hole system, (and  this paper is not devoted to test GR black hole effects, neither horizon nor baryonic effects). The values of the relevant radii here (besides the halo radius): the quantum galaxy radius $r_A$, the BH influence radius $r_i$, and the horizon black hole radius $r_{BH}^{Schw}$ are given by eqs.(3.8)-(3.10). The horizon radius is always extremely  small with respect to the other radii. For galaxies with virial masses from $10^{16} M_\odot$ to $10^7 M_\odot$,   
$r_A$ runs between $0.07$ pc to $1.90$ pc  respectively [as shown in sec. III C], while the horizon radius of the central black hole runs from $10^{-4}$ pc to $10^{-8}$ pc for such range of galaxy masses respectively; $r_i$ is larger than $r_A$ : $r_i > r_A >> r_{BH}^{Schw} $.  
The important point in order to account for both the realistic galaxy and their central SMBH masses, their sizes, velocities, pressure profiles, density profiles and the core sizes, is the DM nature: keV WDM  with its quantum  and its relativistic treatement.

Newtonian black holes have many common properties with general relativity black holes, and  most importantly, they both have the same size.  Recall that Newtonian and post-Newtonian approximation have proven to be remarkably effective even in describing strong-field systems, astrophysical black hole  systems (eg binary bhs) inspiraling towards a final merger, (eg Ref \cite{clifford} and refs therein). Of course, a fully GR treatment is needed to account for a causal space-time structure, central classical space-time curvature singularity, and precise tests of GR, of the horizon or of the "no hair theorems ", for which inner orbits at milliparcec (mpc) distances need to be considered but not for the magnitudes of the galactic masses, sizes, and of their central SMBHs. The GR treatment minimally affects the obtained huge mass magnitude values. A  high merit of the keV WDM approach is that it accounts naturally (with Dark Matter only)  for the realistic astrophysical masses, sizes, density and velocity profiles, rotation curves, equation of state and structural properties of both galaxies and their central SMBHs.}

\end{itemize}

\section{Appendix. Analytic evaluation of the density and the pressure}

The density and the pressure were expressed in sec.
\ref{fortoy} in terms of the integrals
\bea\label{I2I4}
&& I_2(\nu) = 3 \; \int_0^{\infty} y^2 \; dy \; \sqrt{1 +\frac{2 \, y^2}{\tau_1}} \; 
\Psi_{FD}\left(\displaystyle \tau \left[\sqrt{1 +\frac{2 \, y^2}{\tau_1}}-1\right] -\nu\right) \;, \quad
\tau_1 \equiv \frac{m}{T_0}  \; , \cr  \cr  \cr 
&& I_4(\nu) = 5 \; \int_0^{\infty} \frac{y^4 \; dy}{\sqrt{1 +\displaystyle \frac{2 \, y^2}{\tau_1}}} \; 
\Psi_{FD}\left(\displaystyle \tau \left[\sqrt{1 +\frac{2 \, y^2}{\tau_1}}-1\right] -\nu\right) \; .
\eea
We evaluate in this Appendix the integrals $ I_2(\nu) $ and
$ I_4(\nu) $ in the limits $ \nu \gg 1 $ and $ \nu \ll -1 $
corresponding to the quantum and classical regimes,
respectively.

\subsection{The quantum (high density) regime}

In order to evaluate the integrals eqs.(\ref{I2I4}) in the  $ \nu \gg 1 $ regime it is convenient to change
the integration variable $ y $ into $ z $ defined as
\be\label{defvz}
z \;\equiv \; \tau \left[\sqrt{1 +\frac{2 \, y^2}{\tau_1}}-1\right] -\nu \; .
\ee
The density integral $ I_2(\nu) $ takes then the form
\be\label{i2z}
I_2(\nu) = \int_{-\nu}^{\infty} \frac{dz}{e^z + 1} \; h_2(z+\nu) \; , \quad h_2(\nu) \equiv \frac32 \;
\left(\frac{\tau_1}{\tau}\right)^\frac32 \; 
\left(1+ \frac{\nu}{\tau}\right)^2 \; \sqrt{\nu \; \left(1+ \frac{\nu}{2 \; \tau}\right)} \; .
\ee
It is convenient to split the integral eq.(\ref{i2z}) into two pieces
$$
\int_{-\nu}^{\infty}  = \int_{-\nu}^0 + \int_0^{\infty} \; .
$$
Eq.(\ref{i2z}) can then be recasted as
\be\label{i2exa}
I_2(\nu) = \int_0^{\nu} h_2(z) \; dz + \int_0^{\infty} \frac{dz}{e^z + 1} \;\left[ h_2(\nu+z) - h_2(\nu-z) \right] \; ,
\ee
where small terms of the order $ e^{-\nu} $ have been neglected.

{\vskip 0.2cm}

The integral in the first term of eq.(\ref{i2exa}) giving the dominant behaviour of $ I_2(\nu) $
for $ \nu \gg 1 $ can be computed in closed form with the result
\be \label{i2domi}
\begin{split}
&\int_0^{\nu} h_2(z) \; dz = \frac38 \; \left(\frac{\tau_1}2\right)^\frac32 \; 
\left(1 + \frac{\nu}{\tau}\right)
\;\left[ 2 \; \left(1 + \frac{\nu}{\tau}\right)^2 - 1 \right] \;
 \sqrt{\frac{\nu}{\tau} \; \left(2+ \frac{\nu}{\tau}\right)} \; - \\ \\
& - \;  \frac38 \; \left(\frac{\tau_1}2\right)^\frac32 \;      \ln\left[ 1 + \frac{\nu}{\tau} 
+ \sqrt{\frac{\nu}{\tau} \; \left(2+ \frac{\nu}{\tau}\right)} \right].
\end{split}
\ee
Expanding the integrand of the second term of eq.(\ref{i2exa}) in powers of $ z $ and integrating
term by term yields the subdominant terms of $ I_2(\nu) $ for $ \nu \gg 1 $ as an expansion
in inverse powers of $ \nu $
\be\label{i2sub}
\begin{split}
& \int_0^{\infty} \frac{dz}{e^z + 1} \;\left[ h_2(\nu+z) - h_2(\nu-z) \right]  = \\ \\
& = \int_0^{\infty} \frac{dz}{e^z + 1} \;\left[ 
2 \; z \; h_2'(\nu) + \frac{z^3}3 \; h_2'''(\nu) + {\cal O}(z^5) \right] \; = \; \frac{\pi^2}6 \;  h_2'(\nu) \;+\;
\frac{7 \; \pi^4}{360} \; h_2'''(\nu) \;+ \ldots
\end{split}
\ee
We finally get from eqs.(\ref{i2exa})-(\ref{i2sub})
\be 
\begin{split}
&I_2(\nu)\buildrel_{\nu \gg 1}\over= \frac38 \; \left(\frac{\tau_1}2\right)^\frac32 \;
\left(1 + q \right) \;\left[\; 2 \; \left(1 + q\right)^2 - 1 \;\right] \;  \sqrt{q\; \left(2+ q\right)} \; \;-  \\  \\  
&- \;\frac38 \; \left(\frac{\tau_1}2\right)^\frac32 \;   \ln \left[ \; 1 + q + \sqrt{q \; \left(2+ q\right)}  \;\right]\; - \; \frac{\pi^2}{8 \; \sqrt{\nu}} \; \left(\frac{\tau_1}{\tau}\right)^\frac32 \; \frac{\left(1+q\right) \; \left(3 \; q^2 + 6 \; q + 1\right)}{\sqrt{1 + \displaystyle\frac{q}2 }} \; \; + \\ \\
& + \frac{7 \; \pi^4}{640 \; \nu^\frac52} \; \left(\frac{\tau_1}{\tau}\right)^\frac32 \; 
\left(2 \; q^4 + 8 \; q^3 + 7  \; q^2 - 2 \; q + 1\right) \; \frac{(1+q)}{\left(1 + \displaystyle \frac{q}2\right)^\frac52}
\;\; , \quad q \;\equiv \;\frac{\nu}{\tau} \nonumber
\end{split}
\ee
where we used eq.(\ref{i2z}).

\bigskip

The pressure integral $ I_4(\nu) $ can be treated analogously to the density integral $ I_2(\nu) $
using the integration variable $ z $ eq.(\ref{defvz}) 
\be\label{i4z}
I_4(\nu) = \int_{-\nu}^{\infty} \frac{dz}{e^z + 1} \; h_4(z+\nu) \quad  , \quad h_4(\nu) \equiv \frac52 \;
\left(\frac{\tau_1}{\tau}\right)^\frac32 \; \left[\;\nu \; \left(1+ \frac{\nu}{2 \; \tau}\right)\;\right]^\frac32 \; .
\ee
Proceeding as above for $ I_2(\nu) $ we obtain
\be\label{i4exa}
I_4(\nu) = \int_0^{\nu} h_4(z) \; dz + \int_0^{\infty} \frac{dz}{e^z + 1} \;\left[ h_4(\nu+z) - h_4(\nu-z) \right] \; ,
\ee
where small terms of the order $ e^{-\nu} $ have been neglected and the dominant contribution becomes
\be\label{i4domi}
\begin{split}
&\int_0^{\nu} h_4(z) \; dz = \frac5{32 \; \sqrt2} \; \tau \; \tau_1^\frac32 \;
\left [ (2 \; q^2 + 4 \; q -3 ) \; (q + 1) \; \sqrt{q \; \left(q + 2\right)} \right ] \; + \\ \\
&+ \; \frac{15}{32 \; \sqrt2} \; \tau \; \tau_1^\frac32 \; \; \ln \left[ (1 + q)^2 + \sqrt{q \; \left(2+q\right)} \; \right] \;,  \quad   q \; \equiv \; \frac{\nu}{\tau} .  
\end{split}
\ee

The subdominant terms of $ I_4(\nu) $ for $ \nu \gg 1 $ follow by expanding as in eq.(\ref{i2sub}) 
and we finally get from eqs.(\ref{i4exa})-(\ref{i4domi})

\be
\begin{split}
&I_4 (\nu)\buildrel_{\nu \gg 1}\over =  \frac5{32 \; \sqrt2} \; \tau \; \tau_1^\frac32 \;
\left[\; (2 \; q^2 + 4 \; q -3 ) \; (q + 1) \; \sqrt{q \; \left(q + 2\right)} \;\right] \; + \\ \\
&+ \;\frac{15}{32 \; \sqrt2} \; \tau \; \tau_1^\frac32 \;
\; \ln \left[\; (1 + q)^2 + \sqrt{q \; \left(2+q\right)} \;\right] \; + \; \frac{5 \; \pi^2}{8  \; \sqrt2} \; \left(\frac{\tau_1}{\tau}\right)^\frac32 \; 
\sqrt{\nu \; (q+2)} \; (q + 1) \; + \\ \\
& + \;\frac{7 \; \pi^4}{96 \; \sqrt2} \; \left(\frac{\tau_1}{\tau}\right)^\frac32 \; \frac{(\;q + 1\;) \; (\;2 \; q^2 + 4 \; q - 1 \;)}{\left[\;\nu \; (q+2)\; \right]^\frac32} \; .
\end{split}
\ee

\subsection{The classical Boltzmann regime}

In the classical Boltzmann regime $ \nu \ll -1 $ and because  $ e^{\nu } \ll 1 $, the Fermi--Dirac distribution can be
approximated by the exponent of the Boltzmann distribution
\be\label{fdbo}
\frac1{ [\;e^{ (\displaystyle \tau \sqrt{1 +\frac{2 \, y^2}{\tau_1}}-\tau -\nu )} + 1 \; ]} \; = \;
e^{\nu + \tau} \; e^{-\displaystyle\tau \sqrt{1 +\displaystyle\frac{2 \, y^2}{\tau_1}} }\; + \; {\cal O}\left(e^{2 \; \nu}\right)
\ee
 Inserting eq.(\ref{fdbo}) into
 eq.(\ref{I2I4}) for $ I_2(\nu) $ yields
\be\label{I2bol}
\begin{split}
&I_2(\nu) \buildrel_{\nu \ll -1}\over= 3 \;
\exp\left(\nu + \tau\right) \;
\int_0^{\infty} y^2 \; dy \; \sqrt{1
+\frac{2 \, y^2}{\tau_1}} \; 
 e^{-\displaystyle\tau \sqrt{1 +
 \displaystyle\frac{2 \, y^2}{\tau_1}} } + \;
 {\cal O}\left(e^{2 \; \nu}\right) = \\  
& = \; 3 \; \left(\frac{\tau_1}8\right)^\frac32
 \; e^{\nu + \tau} \; \left[\; K_4(\tau) - 
 K_0(\tau) \; \right] + \;
{\cal O}\left(e^{2 \; \nu}\right) \; ,
\end{split}          
\ee
where $ K_n(\tau) $ stands for the Bessel
functions of imaginary argument $ n = 2,
\; 4 $. 

\medskip

Because $ \tau \gg 1 $,  eq.(\ref{I2bol})
can be approximated as
\be\label{I2bol2}
I_2(\nu) \buildrel_{\nu \ll -1 \; , \;
\tau \gg 1}\over= \frac34 \; \sqrt{\pi} \;
e^{\nu} \; \left(\frac{\tau_1}{\tau}\right)^\frac32 \; 
\left[1 + \frac{27}{8 \; \tau} + {\cal O}\left(\frac1{\tau^2}\right) \right] + {\cal O}\left(e^{2 \; \nu}\right) \; .
\ee
The pressure integral $ I_4(\nu) $ can be computed analogously by inserting eq.(\ref{fdbo}) into eq.(\ref{I2I4}) for $ I_4(\nu) $
\be\label{I4bol}
\begin{split}
&I_4(\nu) \buildrel_{\nu \ll -1}\over = 5 \; \exp\left(\nu + \tau\right) \; \int_0^{\infty} 
\frac{y^4 \; dy}{\sqrt{1 +\displaystyle\frac{2 \, y^2}{\tau_1}}} \; 
 e^{-\displaystyle\tau \sqrt{1 + \displaystyle\frac{2 \, y^2}{\tau_1}} } + \; {\cal O}\left(e^{2 \; \nu}\right) \; = \\ 
&= \;\frac{15}{4 \; \sqrt2} \;\; \frac{\tau_1^\frac52}{\tau^2} \; \;e^{\nu + \tau} \; K_2(\tau)\; + \; {\cal O}\left(e^{2 \; \nu}\right) \; .
\end{split}            
\ee
For  $ \tau \gg 1 $  we obtain the simpler expression:
 \be\label{I4bol2}
I_4(\nu) \buildrel_{\nu \ll -1 \; , \; \tau \gg 1}\over= \frac{15}8 \; \sqrt{\pi} \; \left(\frac{\tau_1}{\tau}\right)^\frac52 \; 
e^{\nu} \left[1 + \frac{15}{8 \; \tau} + {\cal O}\left(\frac1{\tau^2}\right) \right] + {\cal O}\left(e^{2 \; \nu}\right) \; .
\ee

\end{document}